\documentclass[submission, Phys]{SciPost}
\usepackage{amsmath}
\usepackage{amssymb}
\usepackage{bbold}
\usepackage{bm}
\usepackage{braket}
\usepackage{caption}
\usepackage{ragged2e}
\usepackage{caption}
\usepackage{csquotes}
\usepackage{derivative}
\usepackage{dcolumn}
\usepackage{dsfont}
\usepackage{float}
\usepackage{graphicx}    
\usepackage{lineno}
\usepackage{mathtools}
\usepackage{physics}
\usepackage{slashed}
\usepackage{subcaption}  
\usepackage{tikz}
\usepackage{verbatim}
\usepackage{xcolor}
\usepackage{ragged2e} 
\usepackage[percent]{overpic}

\usetikzlibrary{decorations.pathmorphing}
\usetikzlibrary{calc}
\usepackage{hyperref}

\usepackage{xr} 
\usepackage{booktabs}
\usepackage{nicematrix} 
\definecolor{giallo}{RGB}{255, 255, 0}
\definecolor{rosso}{RGB}{255, 0, 0}
\definecolor{verde}{RGB}{0, 128, 0}
\definecolor{blu}{RGB}{0, 0, 255}
\definecolor{white}{RGB}{255, 255, 255}

\hypersetup{
    colorlinks,
    linkcolor={red!50!black},
    citecolor={blue!50!black},
    urlcolor={blue!80!black}
}

\usepackage[bitstream-charter]{mathdesign}
\DeclareSymbolFont{usualmathcal}{OMS}{cmsy}{m}{n}
\DeclareSymbolFontAlphabet{\mathcal}{usualmathcal}

\DeclareSymbolFont{usualmathcal}{OMS}{cmsy}{m}{n}
\DeclareSymbolFontAlphabet{\mathcal}{usualmathcal}

\begin{document}

\begin{center}{\Large \textbf{Discrete time crystals in disordered anisotropic Heisenberg chains}}\end{center}

\begin{center}
F. Formicola\textsuperscript{1*},
G. Di Bello\textsuperscript{1*},
A. de Candia\textsuperscript{2,3},
G. De Filippis\textsuperscript{2,3},
C. A. Perroni\textsuperscript{2,3},
\end{center}

\begin{center}
{\bf 1} Dip. di Fisica E. Pancini - Università di Napoli Federico II - I-80126 Napoli, Italy
\\
{\bf 2} SPIN-CNR and Dip. di Fisica E. Pancini - Università di Napoli Federico II - I-80126 Napoli, Italy
\\
{\bf 3} INFN, Sezione di Napoli - Complesso Universitario di Monte S. Angelo - I-80126 Napoli, Italy
\\
* Corresponding authors: F. Formicola, francesco.formicola@unina.it and G. Di Bello, grazia.dibello@unina.it
\end{center}

\section*{Abstract}
{\textbf{
A discrete time crystal is an out-of-equilibrium phase of matter characterized by the spontaneous breaking of discrete time-translation symmetry. Using extensive numerical simulations based on matrix-product-state methods, we provide evidence for discrete time-crystalline behavior in strongly disordered spin chains with Heisenberg interactions, including the isotropic point, subject to periodic driving. Starting from a many-body localized regime, we observe that rotations induced by delta kicks produce a pronounced subharmonic response at half the drive frequency in spin observables. We investigate the stability of this response against rotation-angle errors through entanglement entropy, quantum Fisher information, short-range spin correlations, and restricted-control ergotropy. Increasing the rotation error reveals an intermediate dynamical regime separating the time-crystalline and Floquet-localized responses. In this regime, most observables exhibit signatures of weakly correlated dynamics reminiscent of Anderson localization.
}}

\section{Introduction}

Spontaneous symmetry breaking is one of the most fundamental phenomena in physics, as it underlies the emergence of distinct phases of matter such as ferromagnetism and superconductivity \cite{strocchi2020symmetry}. It occurs when the physical state of a system does not share all the symmetries of its Hamiltonian. While familiar examples involve spatial symmetries and can be formulated within equilibrium statistical mechanics, Wilczek proposed in 2012 the possibility of spontaneous time-translation symmetry breaking \cite{wilczek2012}. Persistent spontaneous breaking of continuous time-translation symmetry was subsequently shown to be excluded in thermal equilibrium under broad conditions \cite{watanabe2015absence}, directing attention toward intrinsically nonequilibrium settings.

A natural framework for studying such phenomena is provided by Floquet systems \cite{sambe1973steady,shirley1965solution,zel1967quasienergy,tsuji2023floquet}, namely quantum systems subject to a time-periodic Hamiltonian, $H(t+T)=H(t)$. Floquet systems possess a discrete time-translation symmetry, since the Hamiltonian is invariant under shifts by integer multiples of the driving period, $t\to t+nT$ \cite{khemani2019brief}. When this symmetry is unbroken, observables can synchronize with the drive and inherit its periodicity \cite{lazarides2014periodic}. By contrast, spontaneous breaking of this discrete time-translation symmetry is associated with a robust subharmonic response, in which observables oscillate with a period that is an integer multiple of the driving period.

Discrete time crystals (DTCs) are nonequilibrium phases characterized by spontaneous breaking of discrete time-translation symmetry, together with a rigid subharmonic response and long-range spatiotemporal order \cite{yao2018,von2016absolute,khemani2016phase,else2016floquet,else2020discrete,yao2017discrete,li2025prethermal}. Since their original theoretical proposals, DTCs have been observed in several experimental platforms, including trapped-ion spin chains, disordered solid-state spin ensembles, and quantum processors \cite{zhang2017observation,choi2017observation,mi2022time,frey2022realization}. These developments have established time-crystalline order as an experimentally accessible form of nonequilibrium many-body dynamics.

A generic isolated interacting Floquet system tends to absorb energy from the drive and approach an infinite-temperature state at long times \cite{abanin2015periodically,d2014long}. Many-body localization (MBL) provides a mechanism for suppressing thermalization in strongly disordered interacting systems \cite{gogolin2016equilibration,d2016quantum,Abanin,Nandki,Alet,Sierant,serbyn2014quantum}. A characteristic dynamical signature of interacting MBL is the slow, logarithmic-in-time growth of entanglement entropy \cite{serbyn2013universal,Bardarson,lukin2019probing}, while local ergotropy has recently been proposed as an additional dynamical witness of localized many-body dynamics \cite{formicola2025local}. In periodically driven systems, Floquet-MBL can suppress indefinite heating and thereby stabilize DTC order, allowing robust subharmonic dynamics to persist \cite{khemani2016phase,else2016floquet,yao2017discrete,ponte2015many}.

MBL is not, however, the only mechanism capable of sustaining time-crystalline behavior. DTC order can also persist for parametrically long times in high-frequency prethermal regimes without strong quenched disorder \cite{else2017prethermal}. More recently, Floquet strong Hilbert-space fragmentation has been identified as a distinct disorder-free stabilization mechanism in periodically kicked XXZ chains \cite{tang2026discrete}. These complementary mechanisms emphasize that the stabilization of robust subharmonic dynamics can originate from different combinations of interactions, disorder, and driving protocols.

Most early theoretical constructions of DTCs focused on Ising-type interactions. Extending robust time-crystalline dynamics beyond predominantly Ising-type settings to systems with genuine Heisenberg exchange is less straightforward, and several different strategies have been explored. In quantum-dot spin chains, Heisenberg exchange has been combined with additional pulse sequences or periodically modulated exchange protocols that generate effective Ising-like dynamics \cite{barnes2019stabilization,van2021protecting,qiao2021floquet}. Time-crystalline behavior has also been investigated in Heisenberg chains subject to strong magnetic-field gradients \cite{li2020discrete}, in central-spin geometries \cite{frantzeskakis2023time}, and in periodically driven Heisenberg chains where integrability and dynamical symmetries play a central role \cite{chen2025discrete}. A two-dimensional DTC with anisotropic Heisenberg interactions has recently been realized on superconducting quantum processors and studied using classical numerical simulations \cite{switzer2026realization}.

Against this background, we investigate a complementary setting: a one-dimensional random-field XXZ Heisenberg chain in the strongly disordered regime, driven by a single global spin rotation per Floquet period. In particular, we ask whether robust time-crystalline behavior can persist at the isotropic Heisenberg point without additional interaction-engineering pulses, strong field gradients, or nonstandard lattice geometries. The parameters of the undriven system are chosen within the many-body-localized regime, while periodic delta kicks generate global rotations about the $x$ axis. This protocol allows us to test whether the interplay of disorder, interactions, and a minimal global drive can stabilize a long-lived subharmonic response in a standard Heisenberg chain.

Our numerical approach is based on tensor-network methods, using matrix product states and matrix product operators \cite{perez2006matrix,white1992density,schollwock2011density}. The time evolution is computed with the two-site time-dependent variational principle (2TDVP) \cite{haegeman2011time,haegeman2016unifying}. Our primary diagnostic is the two-time imbalance correlation function and its frequency-domain response. We complement this analysis with entanglement entropy, quantum Fisher information, restricted-control ergotropy, short-range spatial correlations, and indicators of DTC and localized dynamics. Finite-size comparisons are used to assess convergence over the accessible system sizes and time window.

We find long-lived period-doubled imbalance oscillations and a pronounced response at half the drive frequency, including at the isotropic Heisenberg point. We then investigate the stability of this response against rotation-angle errors. At intermediate errors, the DTC response is suppressed while the entanglement entropy and quantum Fisher information approach stationary values and short-range longitudinal correlations become strongly reduced. Taken together, these signatures are consistent with an effectively weakly correlated intermediate dynamical regime whose behavior is reminiscent of Anderson localization.

The remainder of the paper is organized as follows. Section~2 introduces the disordered XXZ Heisenberg chain and the numerical methods. Section~3 presents the main results, first providing evidence for time-crystalline behavior and then analyzing its robustness against imperfections in the drive. Section~4 summarizes our conclusions, while the appendices provide additional results for different system sizes and parameter regimes.

\section{Model and numerical methods}

We consider a one-dimensional disordered spin-$1/2$ XXZ Heisenberg model with nearest-neighbour interactions, which, without drive, can exhibit ergodic, Anderson-localized, and many-body-localized regimes. 
The Hamiltonian for $N$ spins reads as
\begin{equation}
         \hat{H}_{XXZ}=\frac{J_{\perp}}{2} \sum\limits_{i=1}^{N-1} \bigg(\hat{S}_{i}^{+}\hat{S}_{i+1}^{-}+\hat{S}_{i}^{-}\hat{S}_{i+1}^{+} \bigg)+ J_z\sum\limits_{i=1}^{N-1} \hat{S}_{i}^{z}\hat{S}_{i+1}^{z}+\sum\limits_{i=1}^N h_i\,\hat{S}_i^{z}.
     \label{H_spin}
\end{equation}
At each chain site $i$, the spin operator along the z-axis is $\hat{S}_i^{z}=\frac{1}{2}\hat{\sigma}_i^{z}=\frac{1}{2}\big(\ket{\uparrow}\bra{\uparrow}_i-\ket{\downarrow}\bra{\downarrow}_i\big)$ and the ladder operators $\hat{S}_{i}^{-}=\ket{\downarrow}\bra{\uparrow}_i$ and $\hat{S}_{i}^{+}=\hat{S}_{i}^{-}{}^\dagger$. The quantity $J_{\perp}$ is the spin-flip transverse coupling energy, $J_z$ is the longitudinal interaction strength, and $h_i$ are random on-site fields. The latter are independently sampled from a uniform probability distribution of width $2W$, such that
\begin{equation}
\label{eq:hi}
    h_i \in \big[ -W, W \big ].
\end{equation}
All observables are averaged over multiple independent disorder replica, each with a different stochastically extracted disorder realization (set of local fields $h_i$). The parameter $R$ denotes the total number of disorder realizations.  
Open boundary conditions are imposed for simplicity on a chain with size $N$. In the following, we will use $J_{\perp}$ as reference energy scale, and we set $\hbar=1$.

Through the Jordan-Wigner transformation, the spin model can be mapped onto a one-dimensional model of disordered spinless fermions with nearest-neighbor density-density interactions, as discussed in Appendix\,\ref{app:Jordan-Wigner}. The two models therefore provide equivalent descriptions of the same dynamics.

The system is periodically driven by a sequence of delta kicks described by the time-dependent Hamiltonian
\begin{equation}
    \hat{H}_d(t) =  (\pi-\epsilon)\sum_n\delta(t-nT)\sum^N_{i=1} \hat{S}^x_i.
\end{equation}
Each kick generates a global rotation along the x-axis of all spins of the chain. At the end of each period, if the angle is $\pi$, a perfect flip of the spin z-components is achieved. $\epsilon$ is a parameter introduced to quantify deviations from a perfect $\pi$ rotation and thus of primary importance to examine the robustness of the DTC phase against imperfections in the driving protocol.   
Finally, the Floquet unitary operator for each period reads:
\begin{equation}
    \hat{U}_F(T)=e^{ -i(\pi-\epsilon)\sum_{i=1}^N \hat{S}_i^x}e^{- i \hat{H}_{XXZ} T}.
\end{equation}

To simulate the dynamics, we perform numerical simulations using MPS techniques to represent the quantum states and operators from ITensor library \cite{fishman2022itensor}. Adopting the 2TDVP \cite{haegeman2011time, haegeman2016unifying}, we compute the time-evolved quantum state. 


Unless otherwise stated, the initial state of the chain al along the paper is chosen to be the Néel state
$\ket{\psi}_N=\ket{\uparrow \hspace{2mm} \downarrow \hspace{2mm}  \uparrow \hspace{2mm} \downarrow \hspace{2mm} \dots \uparrow \hspace{2mm} \downarrow }$, typically used for MBL dynamics in these systems \cite{Abanin, Sierant}. Indeed, Néel state is a highly excited state at finite energy density and it is a product state in the local spin basis, making it suitable for probing localization and thermalization dynamics.

\section{Results}
{\it DTC existence.}

We first consider exact $\pi$ pulses by setting the rotation-angle error to $\epsilon=0$.
The main quantity used to identify signatures of DTC behavior in the Heisenberg chain is the two-time correlation function of the imbalance. The imbalance operator is defined as:
\begin{equation}
  \hat{\mathcal{I}}= \frac{1}{N}\sum^N_{i=1} (-1)^{i+1}\hat{\sigma}^{z}_i.
\end{equation}
Its two-time correlation function in the Heisenberg picture reads:
\begin{equation}
    G_2(0,t)= \matrixel{\psi}{ \hat{\mathcal{I}}(0)\hspace{0.5mm} \hat{\mathcal{I}}(t)}{\psi}.
\end{equation}

In this section, we consider the isotropic Heisenberg chain, with $J_z/J_{\perp}=1$. This choice is particularly significant because it represents a substantial departure from previous works, in which the interaction strength $J_{\perp}$ was kept weak compared with the longitudinal interaction $J_z$ in order to remain close to the Ising-like  \cite{barnes2019stabilization, li2020discrete, chen2025discrete, van2021protecting, qiao2021floquet}. 

\begin{figure}[h]
\centering
\begin{overpic}[width=0.49\columnwidth]{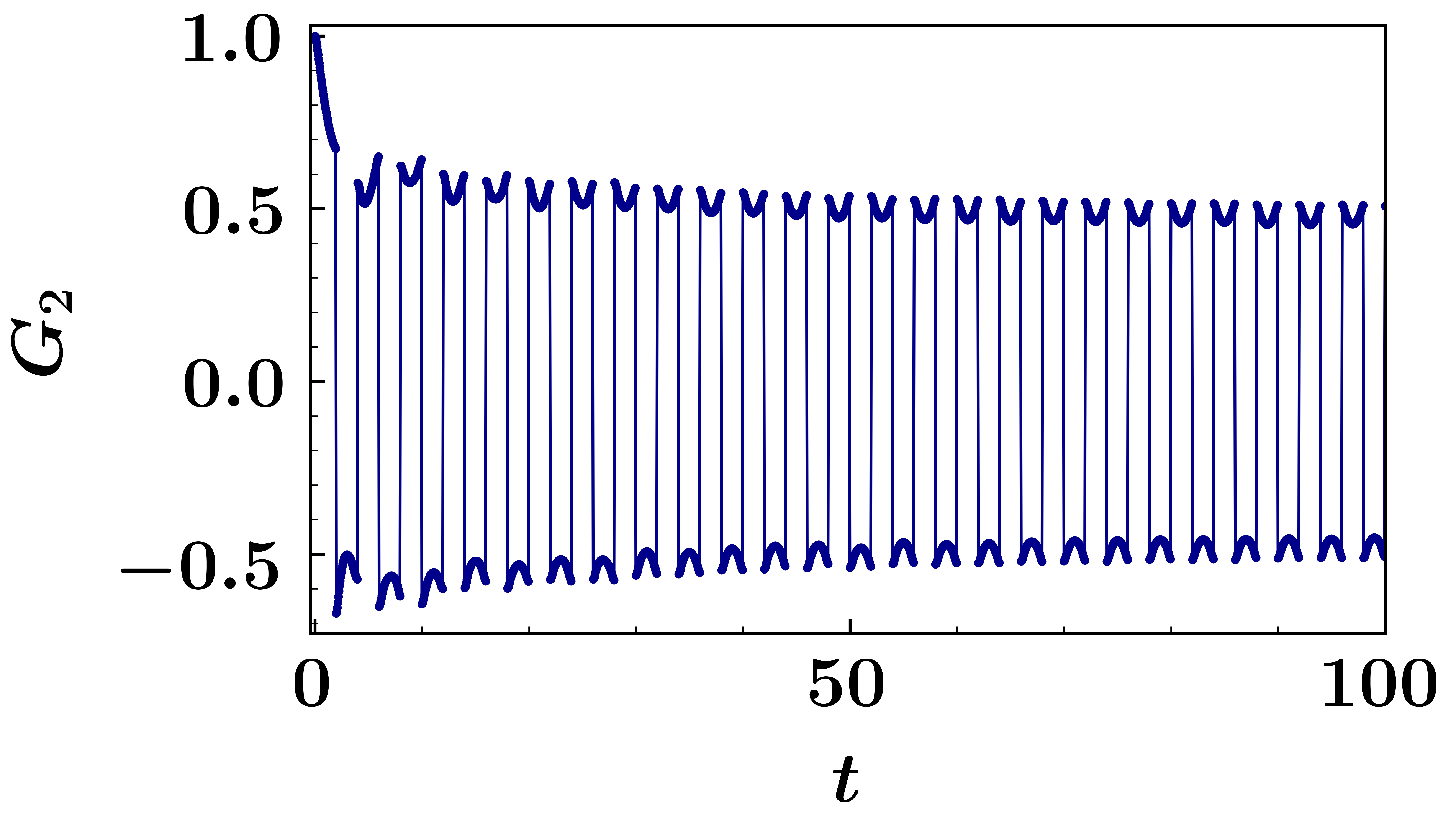}
  \put(-2.5,52){\scalebox{1.5}{\textbf{(a)}}}
\end{overpic}
\hfill
\begin{overpic}[width=0.49\columnwidth]{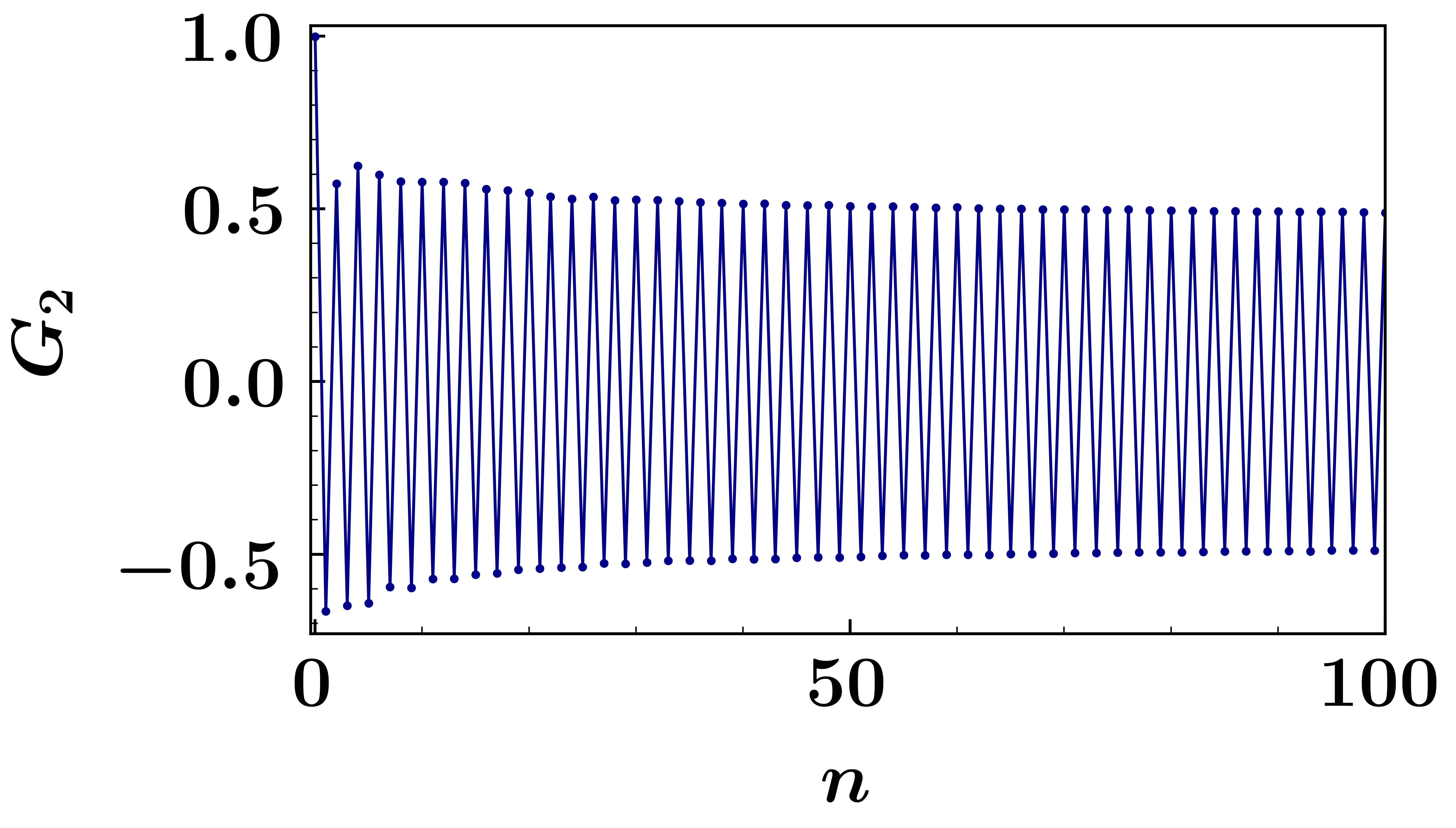}
  \put(-2.5,52){\scalebox{1.5}{\textbf{(b)}}}
\end{overpic}
\begin{overpic}[width=0.5\columnwidth]{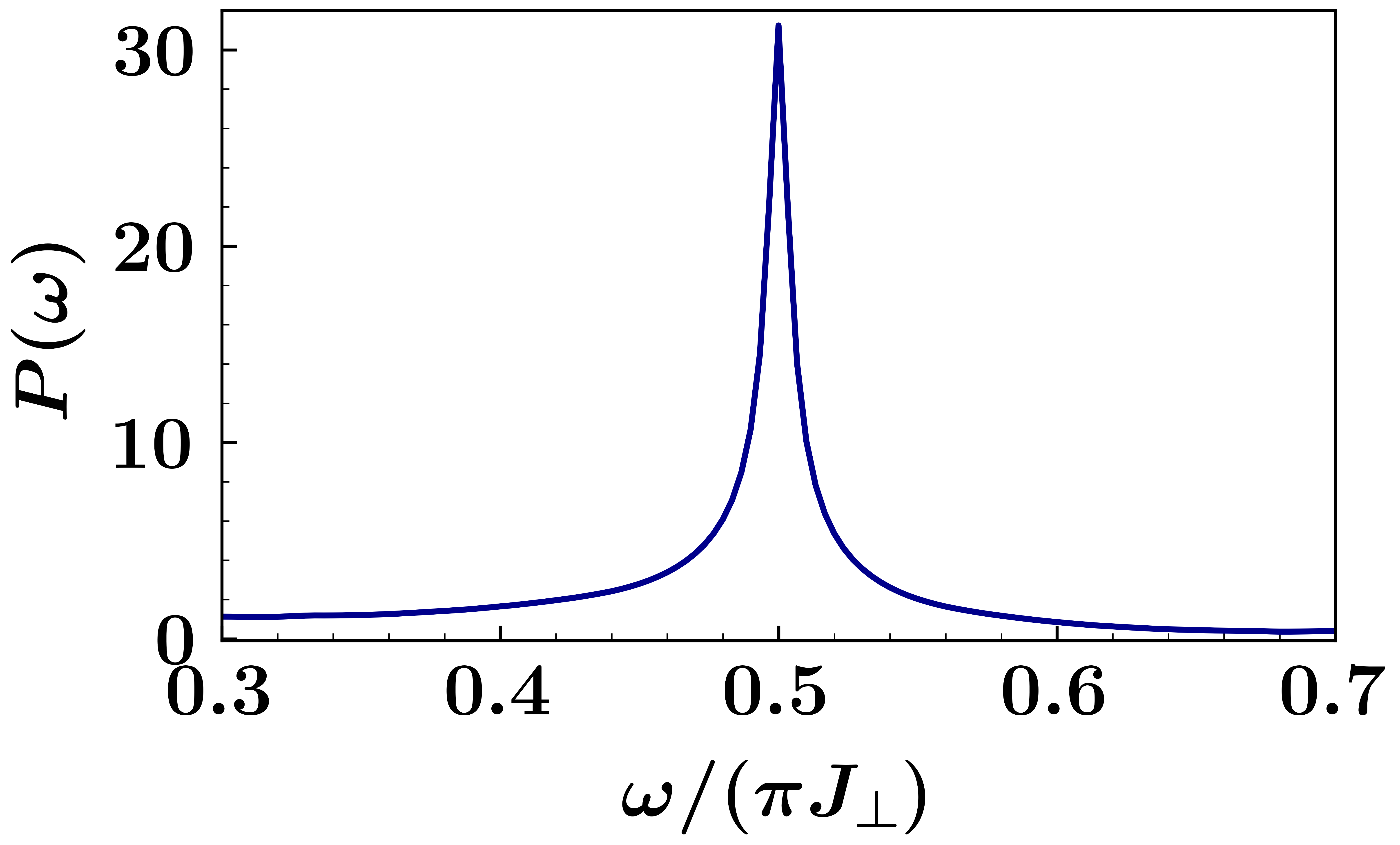}
  \put(-2.5,49){\scalebox{1.5}{\textbf{(c)}}}
\end{overpic}
\caption{\justifying Two-time imbalance correlation function $G_2(0,t)$ as a function of time (a), two-time imbalance correlation function $G_2(0,nT)$ as a function of the number $n$ of Floquet cycles (b) and the spectral analysis of the continuous time signal $G_2(0,t)$ (c). Results are obtained by averaging over $R=5 \cdot 10^3$ independent disorder realizations, interaction strength $J_z/J_{\perp}=1$, disorder strength $W/J_{\perp}=10$, rotation-angle error $\epsilon=0$ with chain length $N=12$ and period $T=2/J_{\perp}$.}
\label{fig:1}
\end{figure}

The upper panels of Fig.\,\ref{fig:1} shows the imbalance correlation function for a driving period $T=2/J_{\perp}$. After a short initial transient, the signal exhibits period-doubled oscillations that remain coherent for more than $100$ Floquet cycles. These results provide evidence for DTC behavior in an isotropic Heisenberg chain driven by a single global rotation during each period.

We also show the spectrum of the $G_2(0,nT)$ in the lower panel of Fig.\,\ref{fig:1} exhibiting a pronounced peak at half the driving frequency, as expected for a DTC: 
\begin{equation}
\omega_{DTC}=\omega_{drive}/2=2\pi/2T=\pi/T.
\end{equation}

The results presented so far were obtained using the Néel state as the initial state. To investigate the dependence of the subharmonic response on the initial condition, we also consider a randomly chosen product state, 
\begin{equation}
\label{psiC}
\ket{\psi}_C=\ket{\uparrow, \hspace{1mm}\downarrow, \hspace{1mm}\downarrow, \hspace{1mm}\uparrow,\hspace{1mm} \downarrow, \hspace{1mm}\dots, \hspace{1mm}\downarrow,\hspace{1mm} \uparrow},
\end{equation}
and an entangled state made by nearest-neighbor singlet pairs: 
\begin{equation}
\label{psiB}
\ket{\psi}_B=\ket{\uparrow,\hspace{1mm} \Psi^-,\hspace{1mm}\Psi^-\dots,\Psi^-,\hspace{1mm}\downarrow},
   \hspace{2cm} \text{where} \hspace{5mm}\ket{\Psi^-}=\frac{\ket{\uparrow}\otimes\ket{\downarrow}-\ket{\downarrow}\otimes\ket{\uparrow}}{\sqrt{2}}.
\end{equation}

\begin{figure}[h]
\centering
\includegraphics[width=0.65\columnwidth]{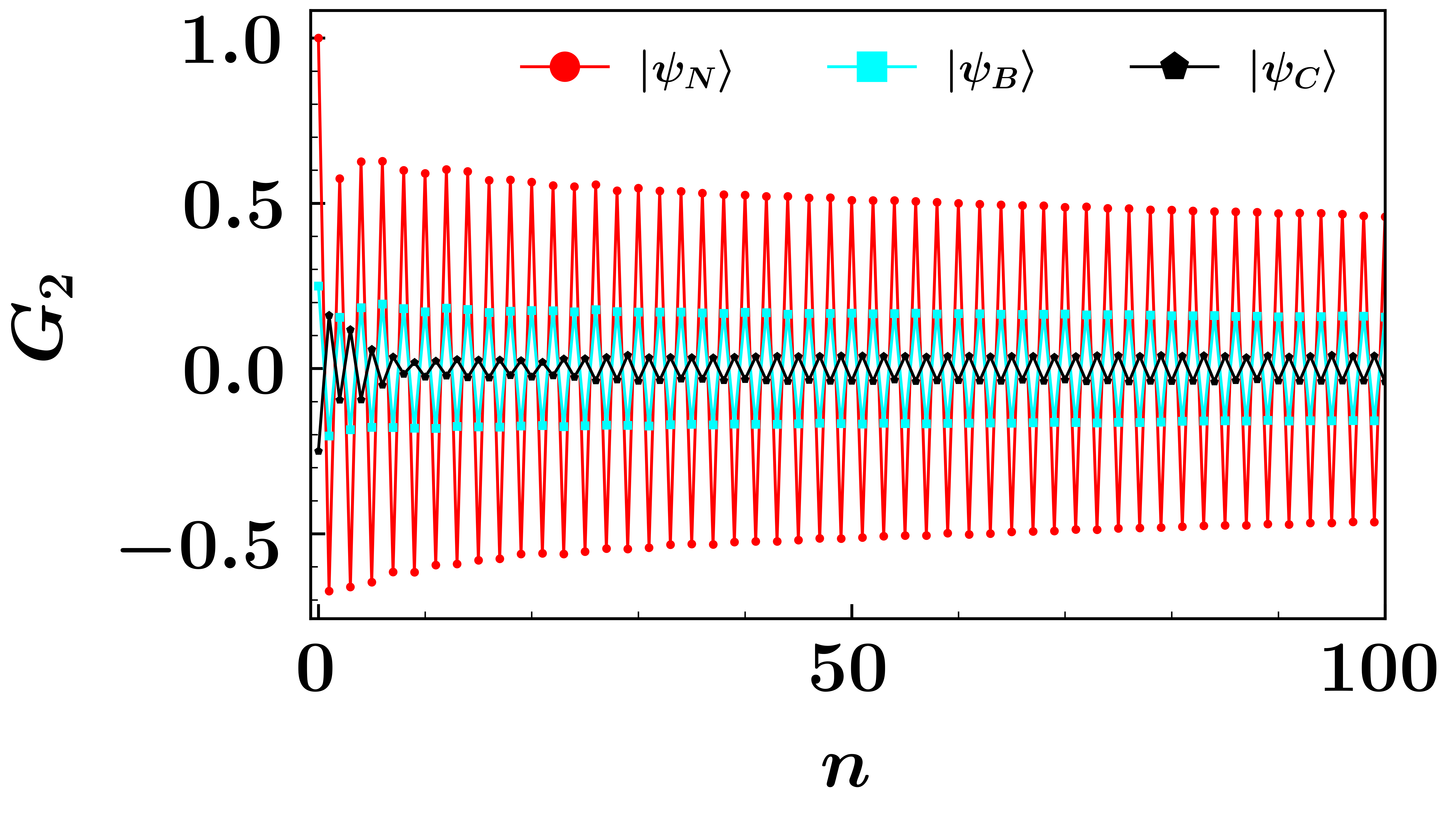}
\caption{\justifying Two-time imbalance correlation function $G_2(0,nT)$ as a function of the number $n$ of Floquet cycles for three different initial states: the Néel state $\ket{\psi}_N$ (red circles), the entangled state $\ket{\psi}_B$ (cyan squares) defined in Eq.\,\eqref{psiB}, and the product state $\ket{\psi}_C$ defined in Eq.\,\eqref{psiC} (black pentagons). Results are obtained by averaging over $R=5 \cdot 10^3$ independent disorder realizations, interaction strength $J_z/J_{\perp}=1$, disorder strength $W/J_{\perp}=10$, rotation-angle error $\epsilon=0$, with chain length $N=8$ and period $T=2/J_{\perp}$.}
\label{fig:2}
\end{figure}

Fig.\,\ref{fig:2} shows that all the initial states considered above exhibit a coherent subharmonic response for the same set of parameters. The amplitude and offset of the oscillations, however, depend strongly on the initial state. Further analysis on the dependence of the period-doubled oscillations on parameters such as the driving period and the longitudinal interaction strength can be found in Appendix\,\ref{app:W} and \ref{app:T}.

So far, we have analysed different initial states, with and without entanglement. If we now consider a state like the following:
\begin{equation}
\ket{\psi}_R=\ket{\phi,\hspace{1mm}\uparrow,\hspace{1mm}\downarrow,\hspace{1mm}\phi,\hspace{1mm}\uparrow,\hspace{1mm}\dots,\hspace{1mm}\uparrow,\hspace{1mm}\phi },
 \hspace{2cm}\text{where} \hspace{5mm}\ket{\phi}=\frac{\ket{\uparrow}+\ket{\downarrow}}{\sqrt{2}},  \label{eq:new_state}
\end{equation}

this is not an eigenstate of the imbalance, so the imbalance correlation function is not simply proportional to the imbalance expectation value. Furthermore, both signals are different from the connected two-points correlation function $G_2^{(c)}(0,t)$:
\begin{equation}
    G_2^{(c)}(0,t)=\matrixel{\psi}{ \hat{\mathcal{I}}(0) \hspace{0.5mm} \hat{\mathcal{I}}(t)}{\psi}-\expval{ \hat{\mathcal{I}}(0)}\expval{ \hat{\mathcal{I}}(t)}.
\end{equation}Even for such state we have recovered long-time coherent period-doubled oscillations, which provide additional evidence for the robustness of the DTC phase.

The dependence of the imbalance correlation function on the chain length $N$ is analysed in Appendix\,\ref{app:N}. We remark there that, for $N=4$ (the case analysed in \cite{switzer2026realization}), the DTC phase is highly unstable. Only when $N$ increases, the phase becomes more stable.

{\it DTC robustness.}

We now investigate the robustness of the DTC regime as a function of the rotation-angle error $\epsilon$. We first present the imbalance correlation function $G_2(0,nT)$ and its frequency spectrum. We then analyse the entanglement entropy, quantum Fisher information, extended ergotropy and two-point spatial correlation functions.

\begin{figure}[]
\centering
  \includegraphics[width=0.7\columnwidth]{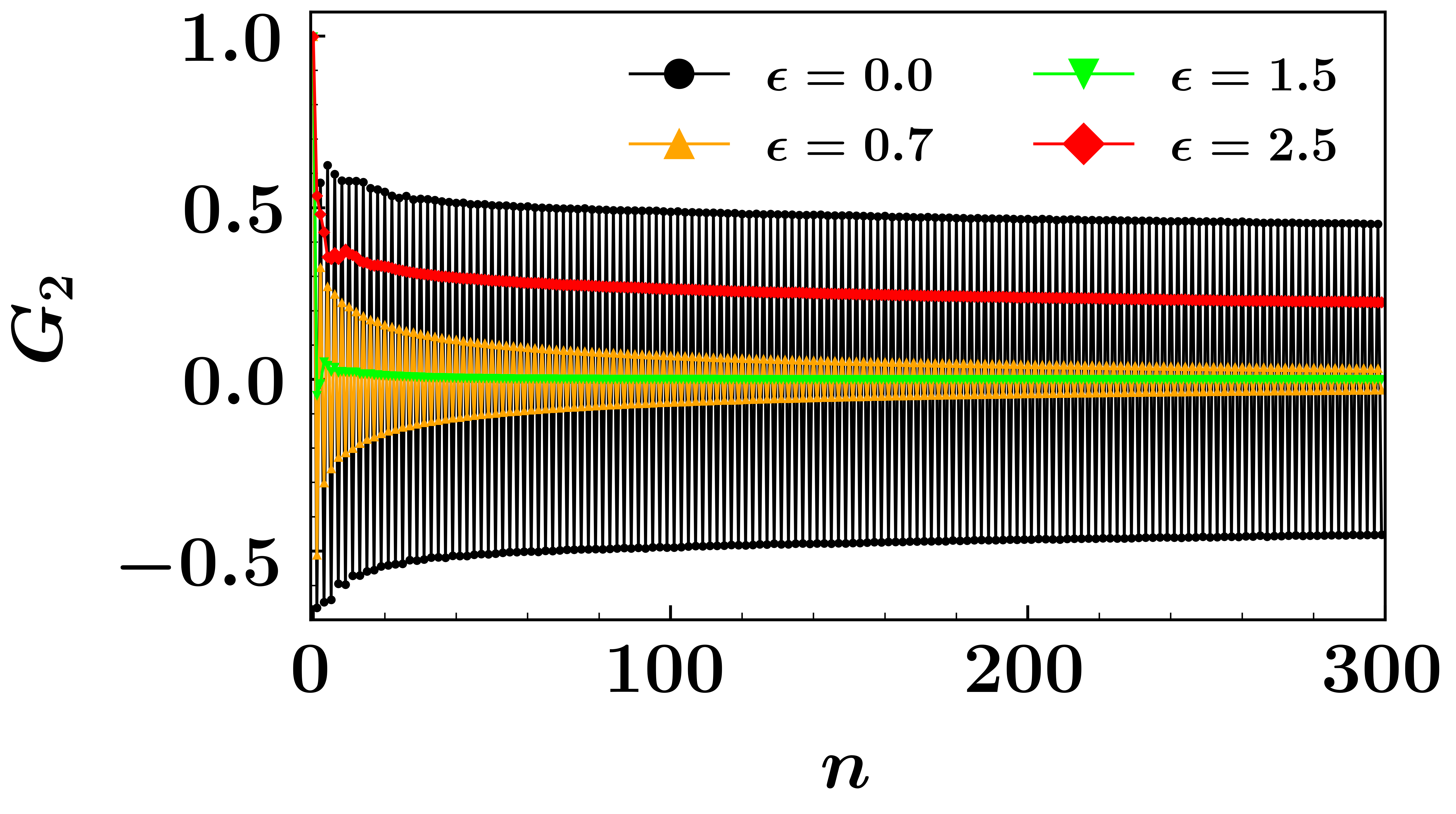}
\caption{\justifying Two-time imbalance correlation function $G_2(0,nT)$ as a function of the number $n$ of Floquet cycles for different values of the rotation-angle error: 
$\epsilon=0.0$ (black circles), $\epsilon=0.7$ (orange upward triangles), $\epsilon=1.5$ (cyan downward triangles) and $\epsilon=2.5$ (lime diamonds). 
Results are obtained by averaging over $R=5 \cdot 10^3$ independent disorder realizations, interaction strength $J_z/J_{\perp}=1$, disorder strength $W/J_{\perp}=10$, with chain length $N=12$ and period $T=2/J_{\perp}$.}
\label{fig:3}
\end{figure}

We examine the imbalance correlation function in Fig.\,\ref{fig:3}. At small rotation errors, the signal exhibits long-lived period-doubled oscillations. Their amplitude progressively decreases as $\epsilon$ increases, up to approximately $\epsilon \simeq 0.7$. After that, as error keeps increasing, the signal becomes essentially flat and the correlation function becomes nearly time independent, remaining close to zero. Finally, for large errors $\epsilon \gtrsim 1.2$, the profile remains flat, but its value becomes nonzero. This behavior is consistent with an MBL regime that retains memory of the initial spin configuration but no longer exhibits a subharmonic response. 

\begin{figure}[]
\centering
  \begin{overpic}[width=0.5\columnwidth]{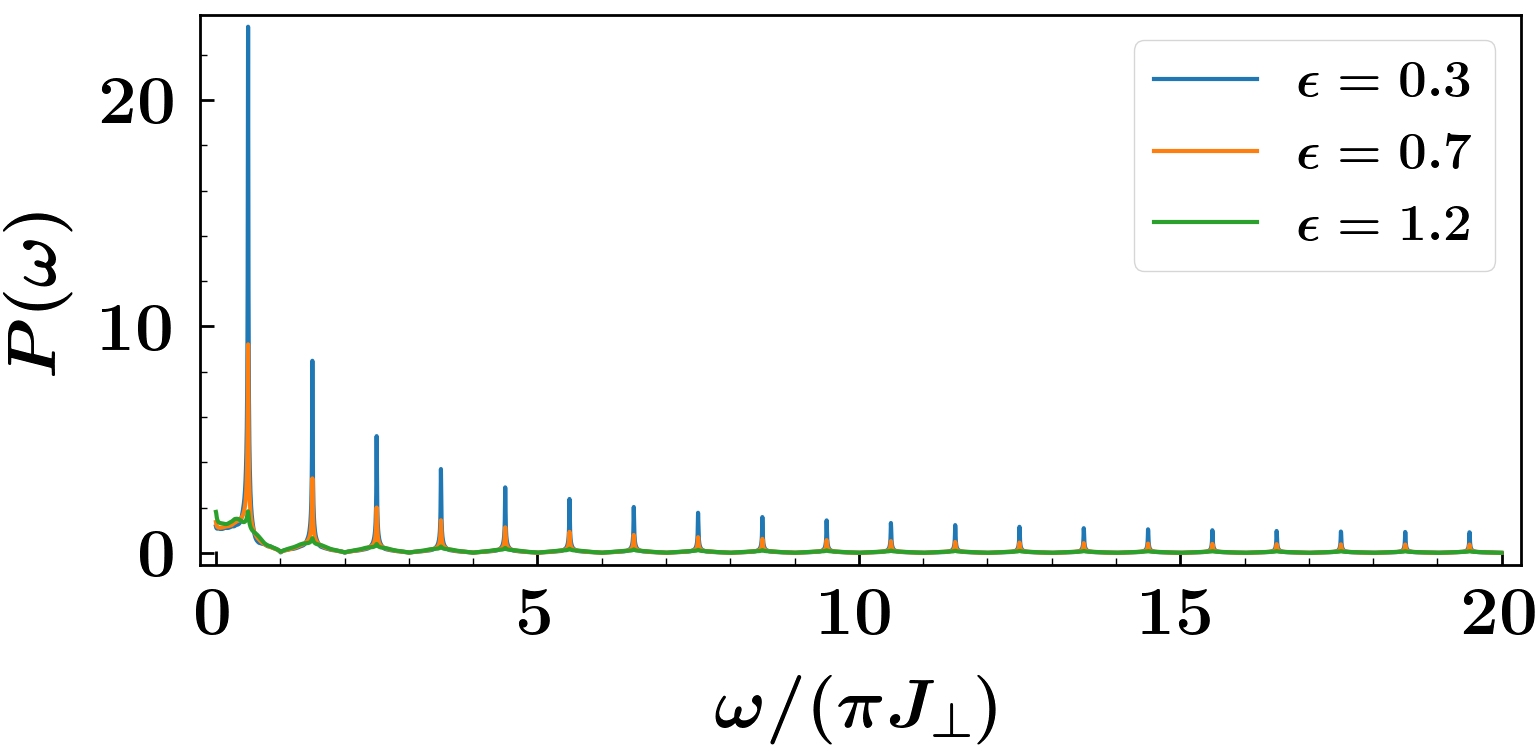}
    \put(-2.8, 46.6){\scalebox{1.5}{\textbf{(a)}}} 
  \end{overpic}
  \hspace{1.7cm}  
  \begin{overpic}[width=0.5\columnwidth]{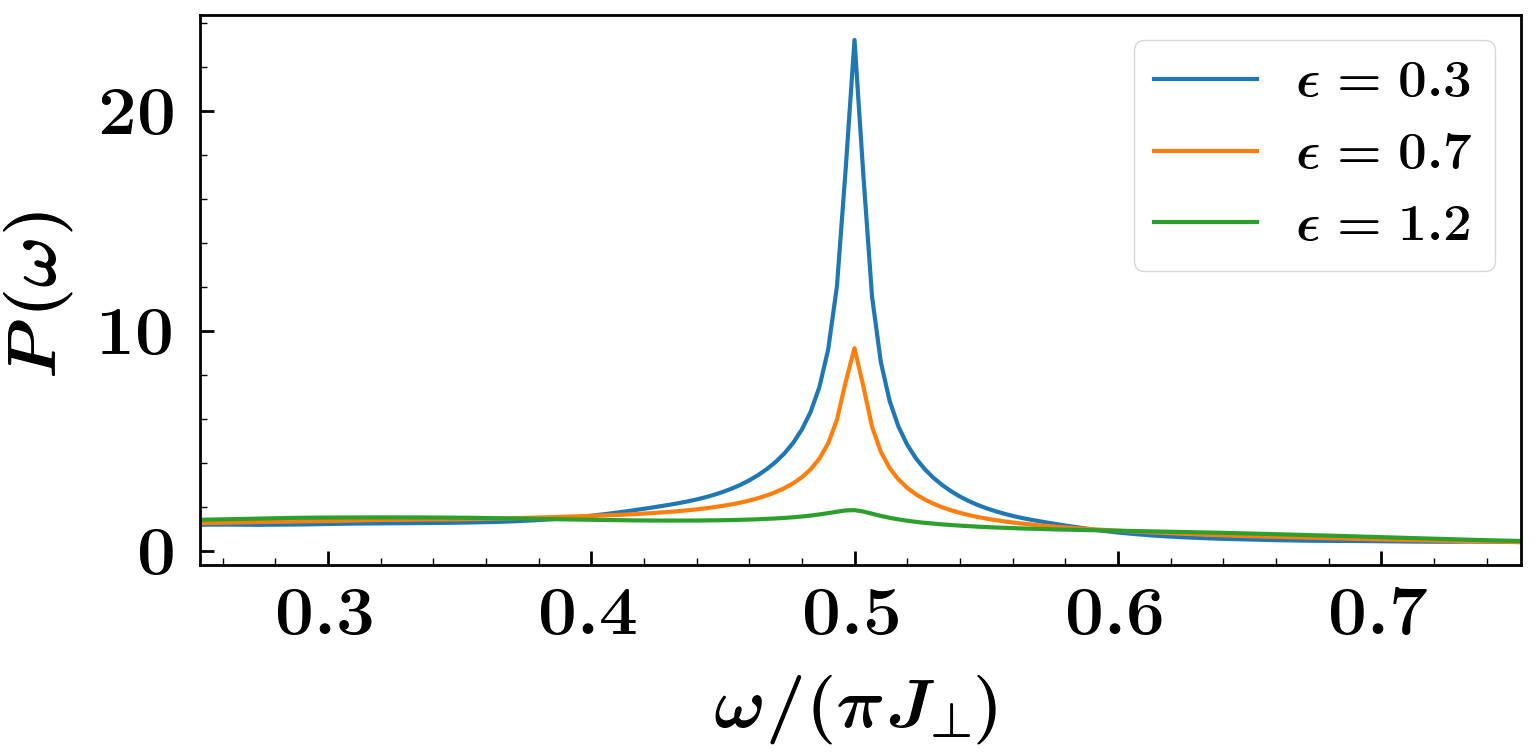}
   \put(-2.8, 46.6){\scalebox{1.5}{\textbf{(b)}}}  
  \end{overpic}
  \hspace{1.7cm}  
  \begin{overpic}[width=0.5\columnwidth]{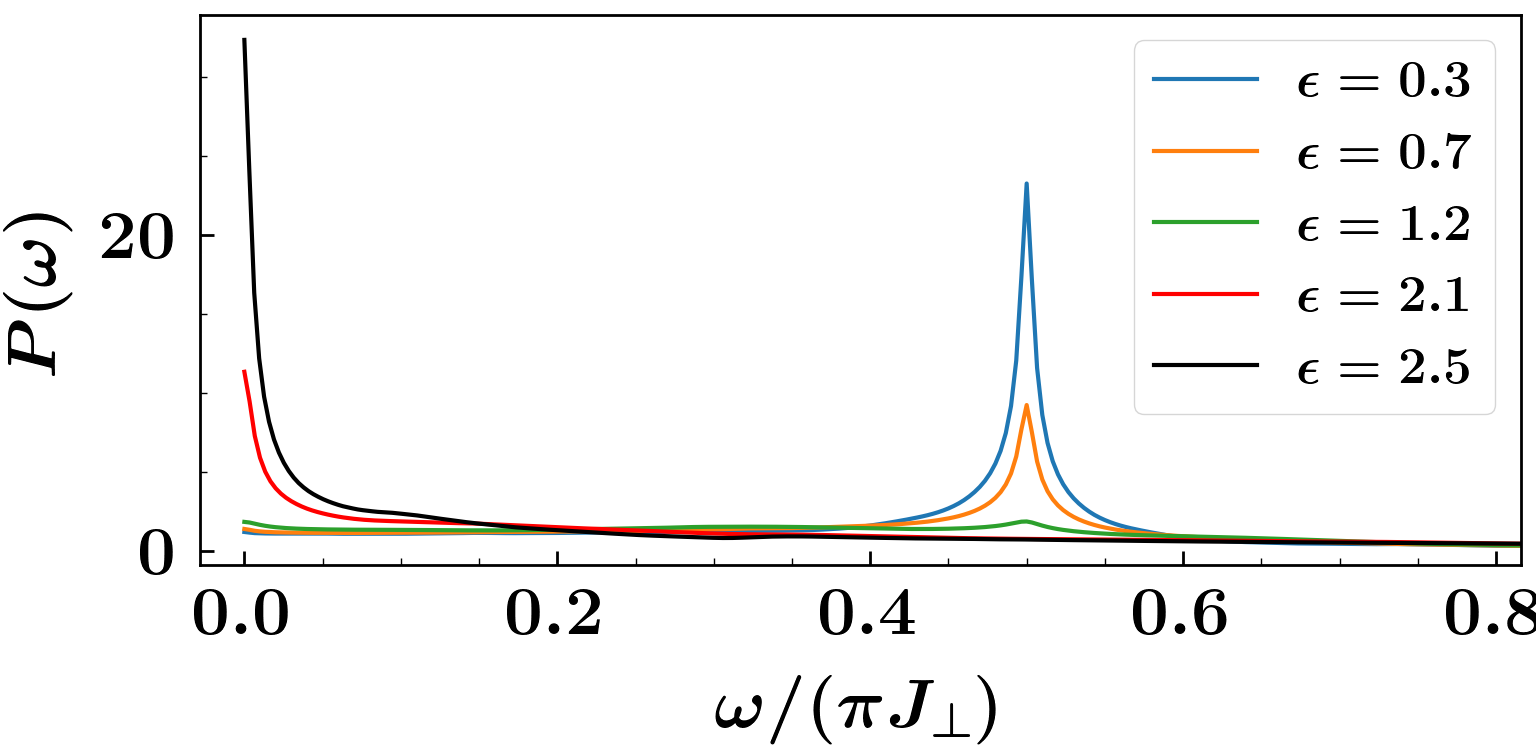}
    \put(-2.8, 46.6){\scalebox{1.5}{\textbf{(c)}}} 
  \end{overpic}
   \caption{\justifying (a) Frequency spectrum of the continuous signal $G_2(0,t)$ as a function of dimensionless frequencies for different values of the rotation-angle errors: $\epsilon=0.3$ (blue), $\epsilon=0.7$ (orange), $\epsilon=1.2$ (green). (b) Magnification of the spectrum around the fundamental subharmonic peak at $\omega_{\mathrm{DTC}}=\pi/T$. (c) Magnification of the low-frequency region around $\omega=0$, presenting also the cases $\epsilon=\{1.7,\hspace{1mm}  2.1,\hspace{1mm} 2.5\}$. Results are obtained by averaging over $R=5 \cdot 10^3$ independent disorder realizations, disorder strength $W/J_{\perp}=10$, with chain length $N=12$ and period $T=2/J_{\perp}$.}
\label{fig:4}
\end{figure}

The progressive breakdown of the DTC response is further illustrated in Fig.\,\ref{fig:4} by the frequency-domain analysis of the continuous signal of two-time correlation function $G_2(0,t)$. For small errors, the spectrum displays a pronounced peak at the subharmonic frequency $\omega_{DTC}=\pi/T$ (see Fig.\,\ref{fig:4} (a)). Additional peaks occur at odd harmonics of the fundamental DTC frequency, $\omega_m=(2m+1)\frac{\pi}{T}$, for $m\in \mathbb{N}_0$. As the rotation error increases, the subharmonic peak loses spectral weight and eventually disappears, as shown in Fig.\,\ref{fig:4} (b). At the same time, the spectral weight near zero frequency increases (see Fig.\,\ref{fig:4} (c)), consistently with the increasingly stationary behavior of the imbalance correlation function after the breakdown of the period-doubled response.

We next introduce the entanglement entropy. Given a system at time $t$ bipartitioned into two subsystems, A and B, entanglement entropy is defined as the von Neumann entropy of one of the subsystems:
\begin{equation}
    S_{ent}(t)=-\Tr\{ \hat{\rho}_A(t)\ln\hat{\rho}_A(t) \}.
\end{equation}
For a globally pure state, the entanglement entropies of the two subsystems coincide. In quantum information, this is a fundamental quantity since it measures how entanglement is generated across the bipartition during the dynamics. It also provides a useful diagnostic of localization: whereas thermalizing systems generally exhibit rapid entanglement growth, an MBL system is characterized by a much slower, typically logarithmic, increase of the entanglement entropy \cite{formicola2025local}. 

Another powerful tool that we can adopt in order to investigate the robustness of the DTC phase is the quantum Fisher information (QFI), which quantifies the sensitivity of a quantum state to changes in an external parameter. It plays a central role in quantum metrology because it determines the ultimate precision allowed in parameter estimation \cite{liu2020quantum,helstrom1969quantum, holevo2011probabilistic,gammelmark2014fisher,alipour2014quantum, parlato2025quantum}. It is also sensitive to multipartite correlations and has been employed as a diagnostic of localized many-body dynamics \cite{ nagao2026probing, de2019algebraic, liu2018quantum}. 
For a state $\rho(\boldsymbol{x})$ depending on a set of parameters $x_{\mu}$, the QFI matrix can be written as 
\begin{equation}
    F_{\mu\nu} = \frac{1}{2}\Tr(\rho\{L_{\mu},L_{\nu}\}),
\end{equation}
where $\{L_{\mu},L_{\nu}\}$ is the anti-commutator of the symmetric logarithmic derivatives $L_{\mu}$ that are solutions of the equation $\pdv{\rho}{x_{\mu}}=\frac{1}{2}(L_{\mu}\rho +\rho L_{\mu})$. 
As pointed out in \cite{switzer2026realization,hauke2016measuring}, we can associate the total magnetization QFI of a spin system to its magnetic susceptibility, the normalized sum of the two-point spatial correlation functions. In the present work, we consider the QFI associated with collective rotations generated by the total magnetization along the z direction $M_z=\frac{1}{2}\sum_{i=1}^{N}\sigma_i^z$. 
The QFI associated with $M_z$ is therefore equal to four times its variance and the QFI per spin reads:
\begin{equation}
        F_Q(t)=\frac{1}{N}\sum_{i,j}C_{ij}(t)=\frac{1}{N}\sum_{i,j} (\expval{\hat{\sigma}^{z}_i\hat{\sigma}^{z}_j}_t-\expval{\hat{\sigma}^{z}_i}_t\expval{\hat{\sigma}^{z}_j}_t)=\chi(t).
    \label{Fq}
\end{equation}
The normalized QFI therefore measures the collective fluctuations of the magnetization. The QFI is evaluated separately for each pure-state disorder realization and subsequently averaged over the disorder ensemble.

We also consider the ergotropy, a central tool in quantum thermodynamics \cite{allahverdyan2004maximal, alicki2013entanglement, SalviaGiovannetti, formicola2025local}. The ergotropy is defined as the maximum amount of work extractable from an isolated quantum system via cyclic unitary transformations.
Inspired by the extended-local-ergotropy framework of
Ref.~\cite{castellano2024extended}, we define a restricted-control ergotropy
adapted to the operations available in the present spin-chain model.
Let $\mathcal U_{\mathrm{ex}}^{\mathrm{fin}}$ be the set of all finite sequences obtained by alternating free evolution under $ H_{\mathrm{XXZ}}$ with admissible global control unitaries,
\begin{equation}
     \hat{V} =
 \hat{U}_{\mathrm c}^{(m)}
e^{-i \hat{H}_{\mathrm{XXZ}}\tau_m}
\cdots
 \hat{U}_{\mathrm c}^{(1)}
e^{-i \hat{H}_{\mathrm{XXZ}}\tau_1},
\qquad
 \hat{U}_{\mathrm c}^{(k)}\in\mathcal C,
\quad
\tau_k\geq0.
\end{equation}
The delta-kick rotations used in the present dynamics are particular elements of the control class $\mathcal C$. We denote by $\overline{\mathcal U}_{\mathrm{ex}}$ the closure of $\mathcal U_{\mathrm{ex}}^{\mathrm{fin}}$.

At stroboscopic times, the energy is evaluated with respect to the static Hamiltonian $H_{\mathrm{XXZ}}$. The restricted-control ergotropy is then
\begin{align}
\mathcal E_{\mathrm{ex}}(t)
=
\sup_{ \hat{V}\in\mathcal U_{\mathrm{ex}}^{\mathrm{fin}}}
\operatorname{Tr}
\left\{
\hat{H}_{\mathrm{XXZ}}
\left[
\hat{\rho}(t)- \hat{V}\hat{\rho}(t) \hat{V}^\dagger
\right]
\right\}
\nonumber
=
\max_{ \hat{V}\in\overline{\mathcal U}_{\mathrm{ex}}}
\operatorname{Tr}
\left\{
 \hat{H}_{\mathrm{XXZ}}
\left[
\hat{\rho}(t)- \hat{V}\hat{\rho}(t) \hat{V}^\dagger
\right]
\right\},
\end{align}
where the second equality follows from the finite dimensionality of the chains considered here.

If
$\rho(t)= U(t)\rho(0) U^\dagger(t)$ with
$ U(t)\in\mathcal U_{\mathrm{ex}}^{\mathrm{fin}}$, compactness implies that $\overline{\mathcal U}_{\mathrm{ex}}$ is a group and hence contains $ U^\dagger(t)$. Therefore,
\begin{equation}
\mathcal E_{\mathrm{ex}}(t)
\geq
\operatorname{Tr}
\left\{
 \hat{H}_{\mathrm{XXZ}}
\left[
\hat{\rho}(t)-\hat{\rho}(0)
\right]
\right\}
=
\left\langle \hat{H}_{\mathrm{XXZ}}\right\rangle_t
-
\left\langle \hat{H}_{\mathrm{XXZ}}\right\rangle_0.
\end{equation}
The inverse transformation may belong only to the closure of the finite accessible sequences and therefore need not correspond to an exact finite-time return protocol.

We finally focus on two indicators, exploited in \cite{switzer2026realization}, to better characterize the MBL and DTC phases:
\begin{equation}
       \Delta_{MBL} = \frac{1}{N_c} \sum_{n=0}^{N_c-1} \abs{\expval{\hat{\mathcal{I}}(nT)}},
        \hspace{2cm}
       \Delta_{DTC} = \frac{1}{N_c} \sum_{n=0}^{N_c-1} (-1)^{n}\expval{\hat{\mathcal{I}}(nT)}.
    \label{DELTA_eq}
\end{equation}
Here, $N_c$ is the total number of Floquet cycles in the time average, and $nT$ is the stroboscopic time at the end of each period. $\Delta_{MBL}$ is meant to be zero in ergodic phase and non-zero in both DTC and MBL phases since it measures the persistence of the initial imbalance. On the other hand, $\Delta_{DTC}$ is supposed to be non-zero only in DTC regime since it selects the component of the imbalance that alternates as $(-1)^n$ and is therefore sensitive to a period-doubled response.

\begin{figure*}[]
    \centering

    \begin{minipage}{0.48\textwidth}
        \centering
        \includegraphics[width=\linewidth]{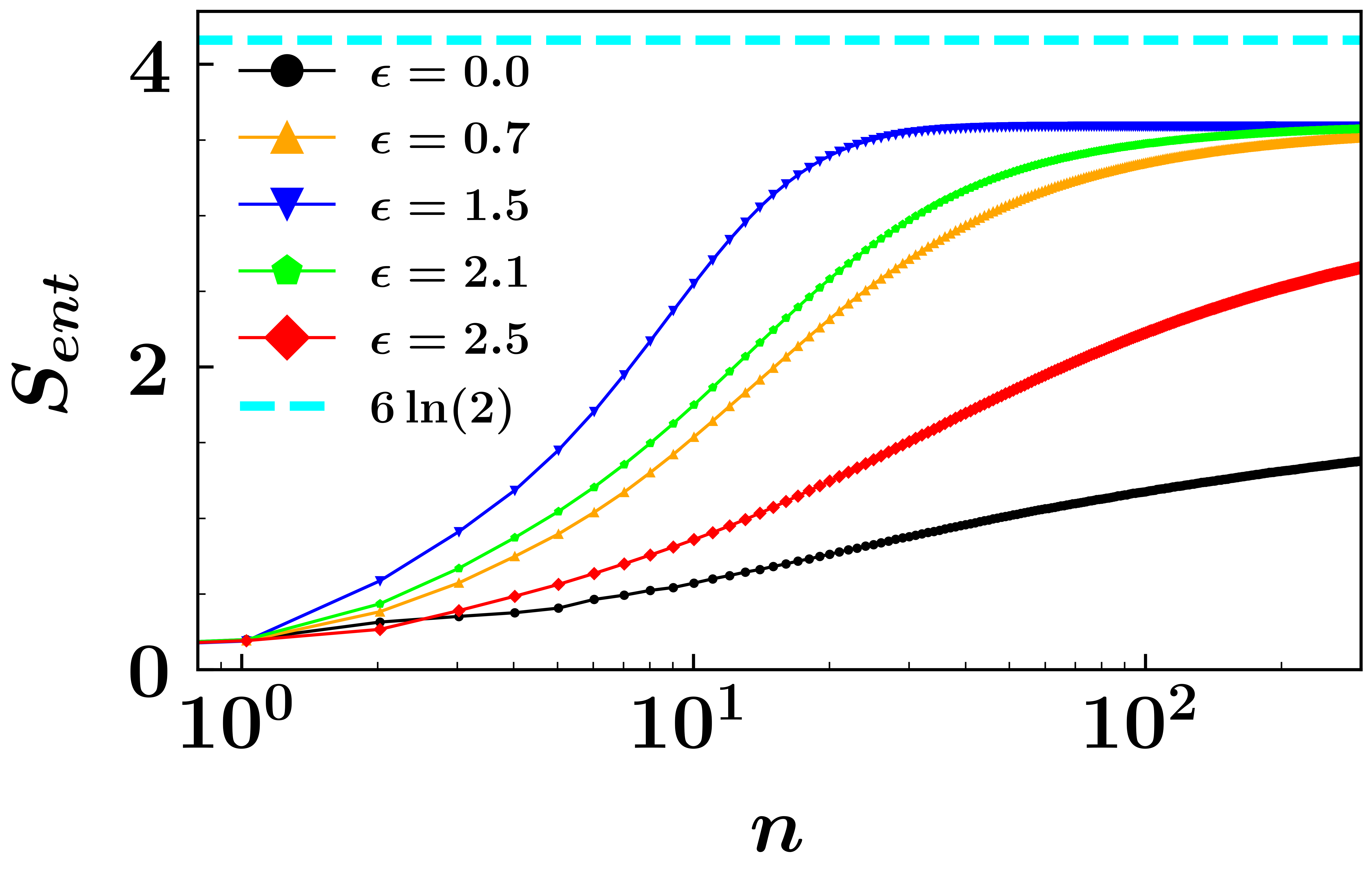}
         \put(-220.5, 120.0){\scalebox{1.5}{\textbf{(a)}}} 
    \end{minipage}
    \hfill
    \begin{minipage}{0.48\textwidth}
        \centering
        \includegraphics[width=\linewidth]{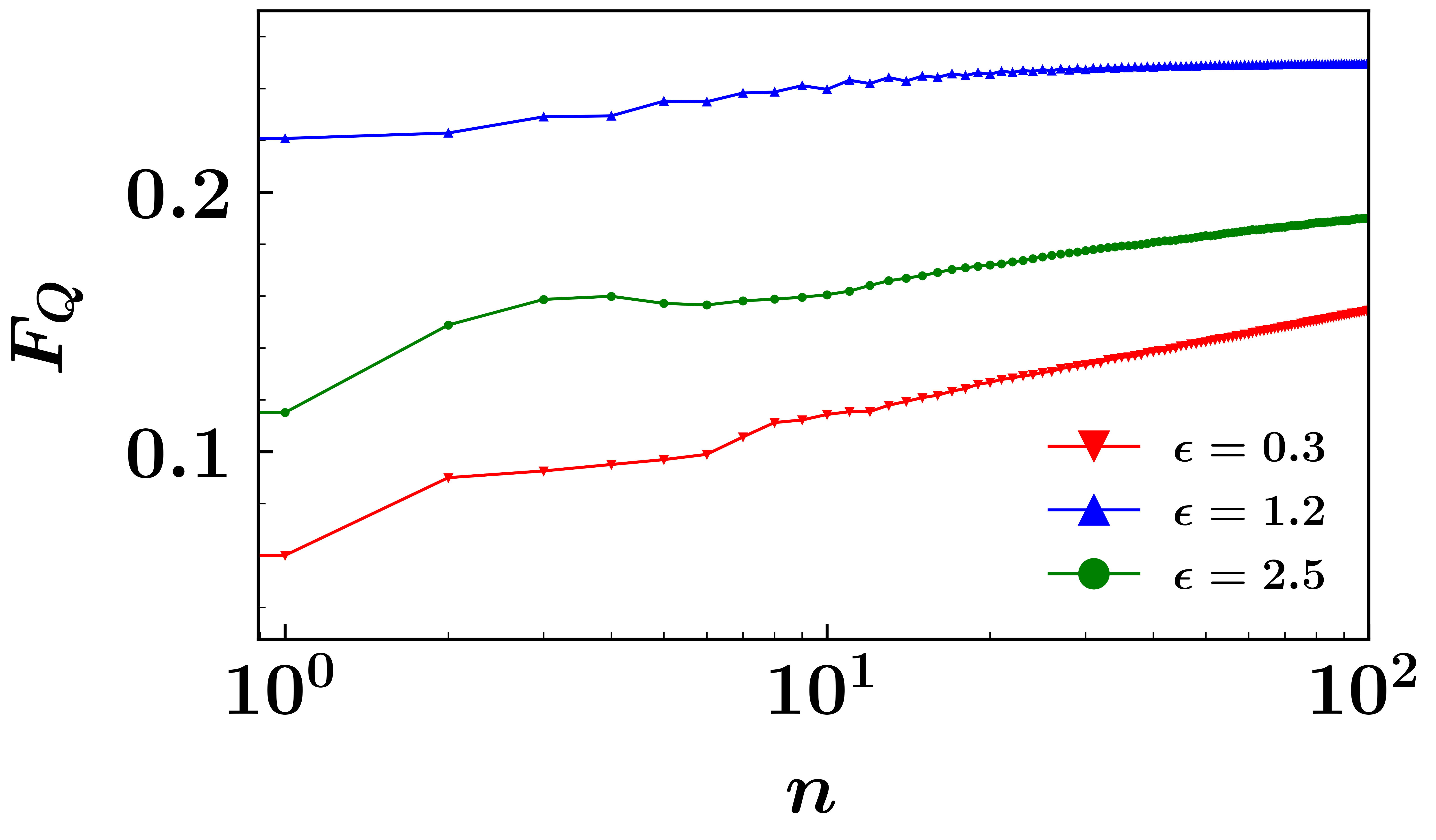}
       \put(-210.5, 120.0){\scalebox{1.5}{\textbf{(b)}}}
    \end{minipage}

    \vspace{0.0cm}

    \begin{minipage}{0.48\textwidth}
        \centering
        \includegraphics[width=\linewidth]{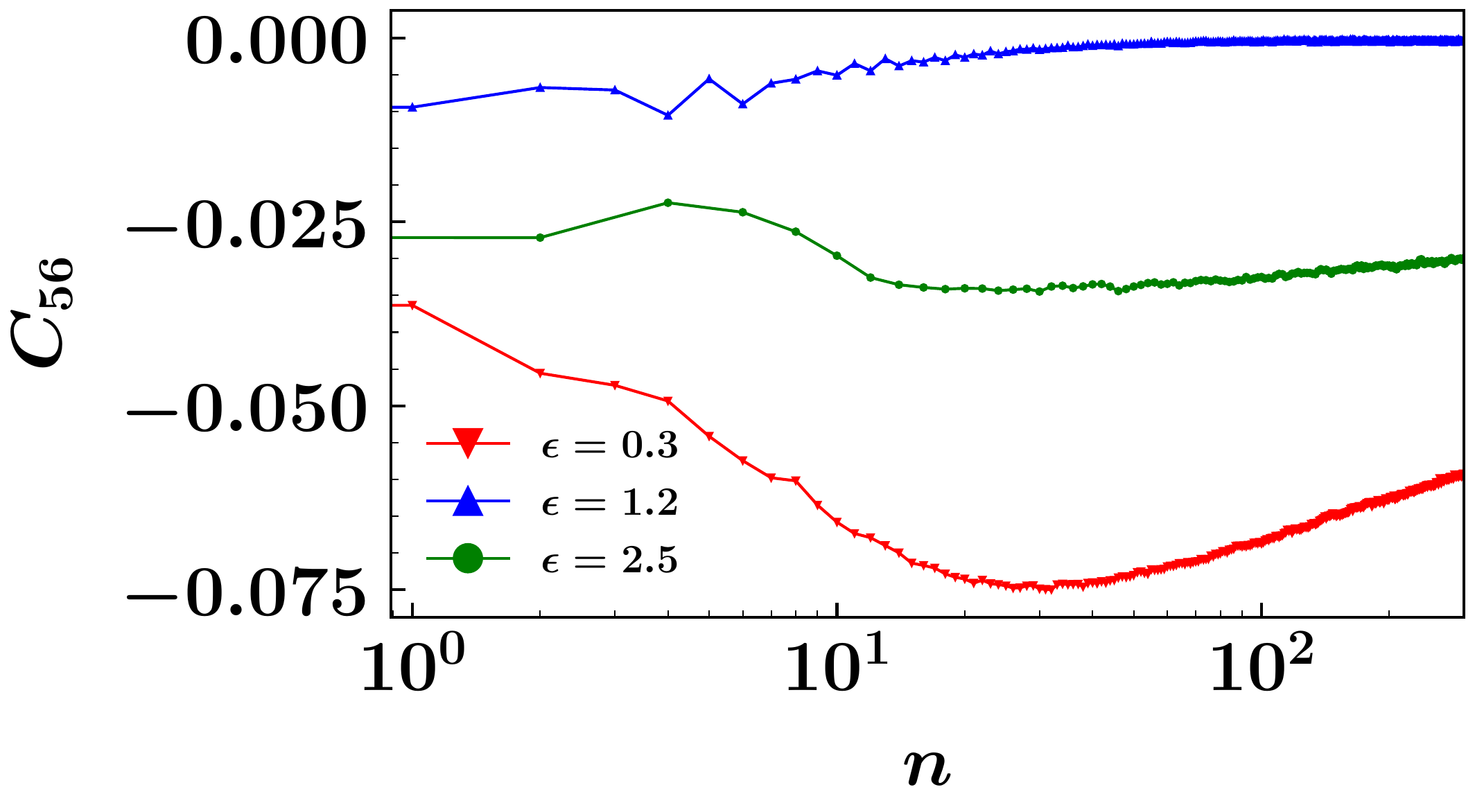}
        \put(-220.5, 120.0){\scalebox{1.5}{\textbf{(c)}}}
    \end{minipage}
    \hfill
    \begin{minipage}{0.48\textwidth}
        \centering
        \includegraphics[width=\linewidth]{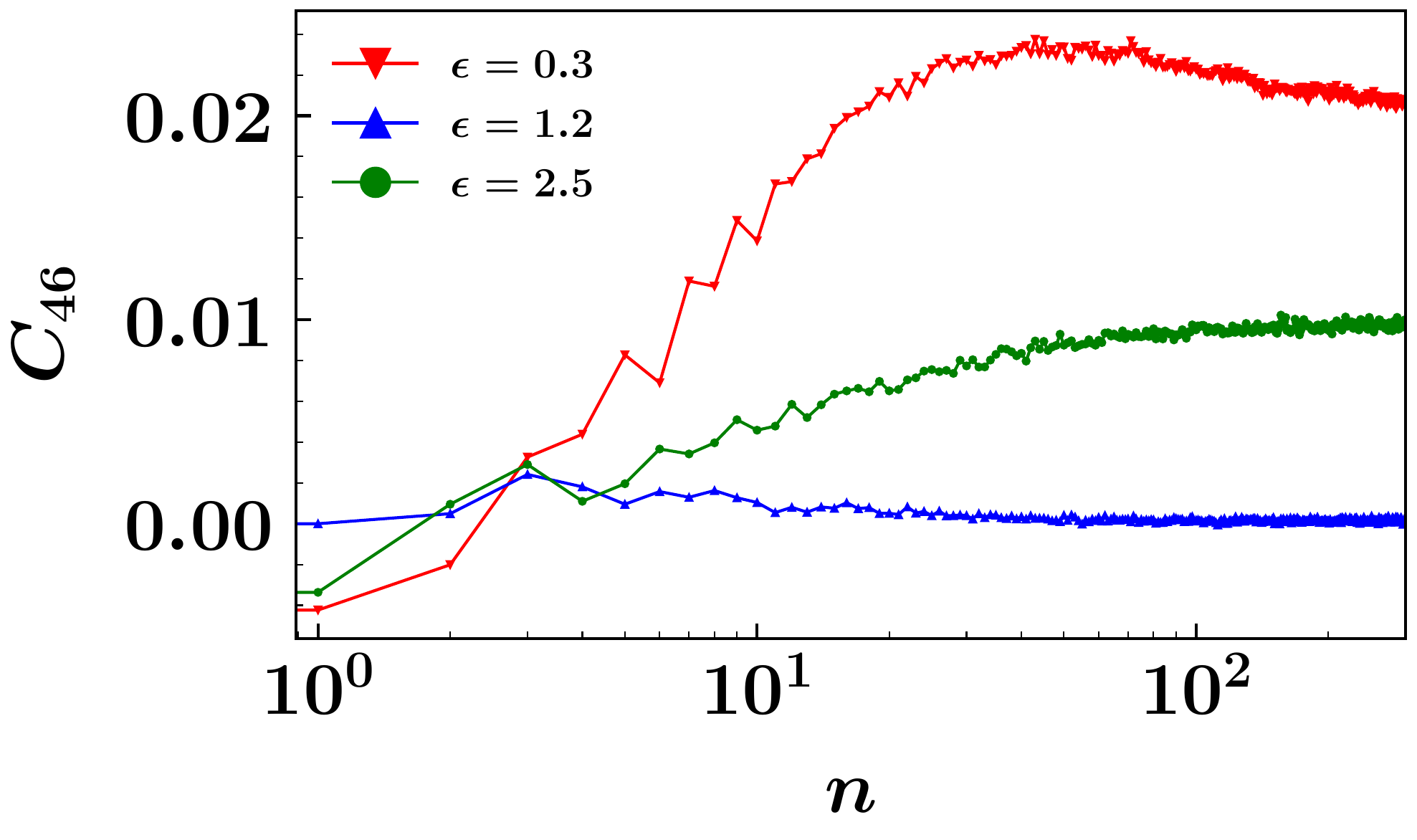}
         \put(-210.5, 120.0){\scalebox{1.5}{\textbf{(d)}}}
    \end{minipage}

 \vspace{0.5cm}

    \begin{minipage}{0.48\textwidth}
        \centering
        \includegraphics[width=\linewidth]{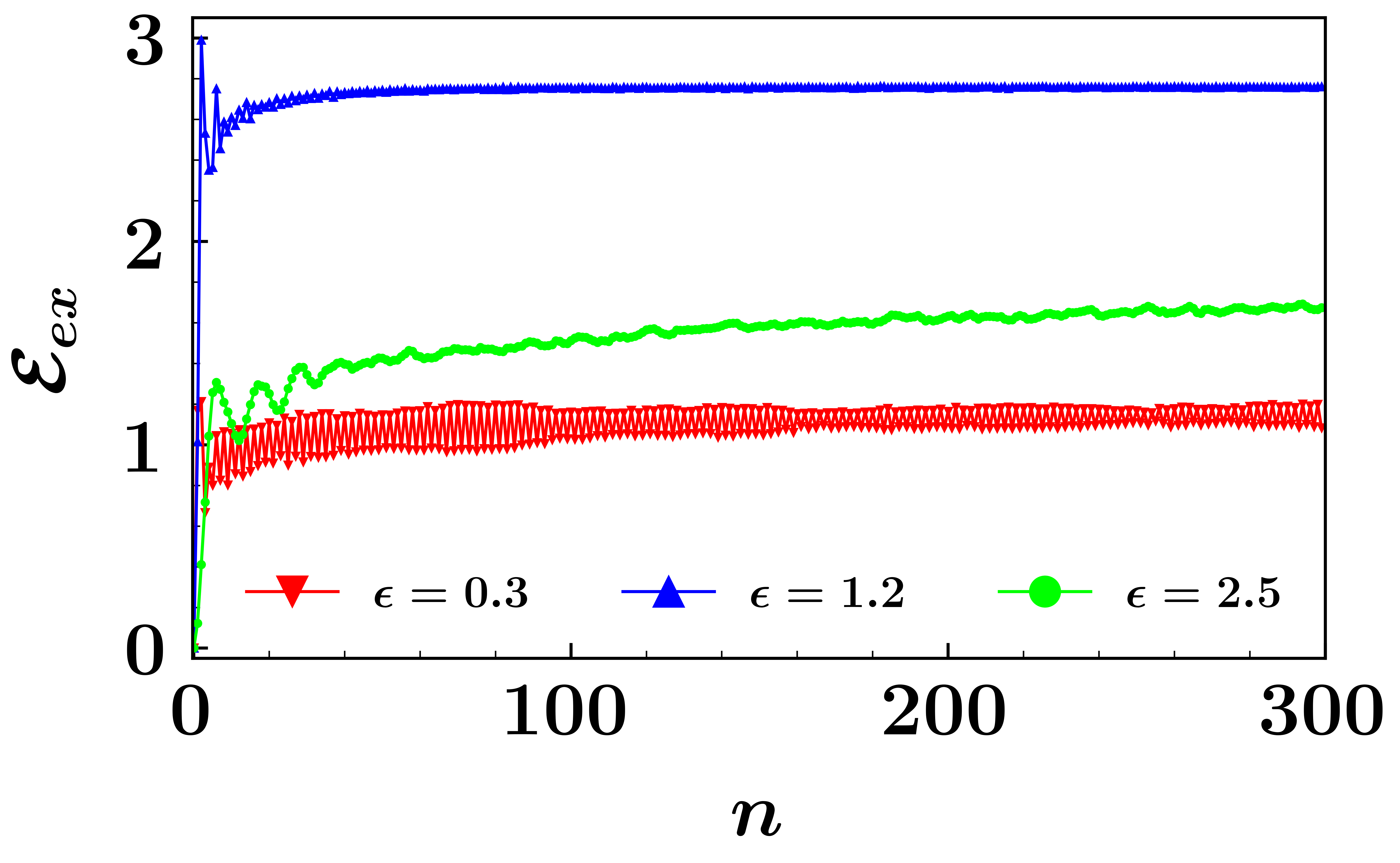}
       \put(-220.5, 120.0){\scalebox{1.5}{\textbf{(e)}}}
    \end{minipage}
    \hfill
    \begin{minipage}{0.48\textwidth}
        \centering
        \includegraphics[width=\linewidth]{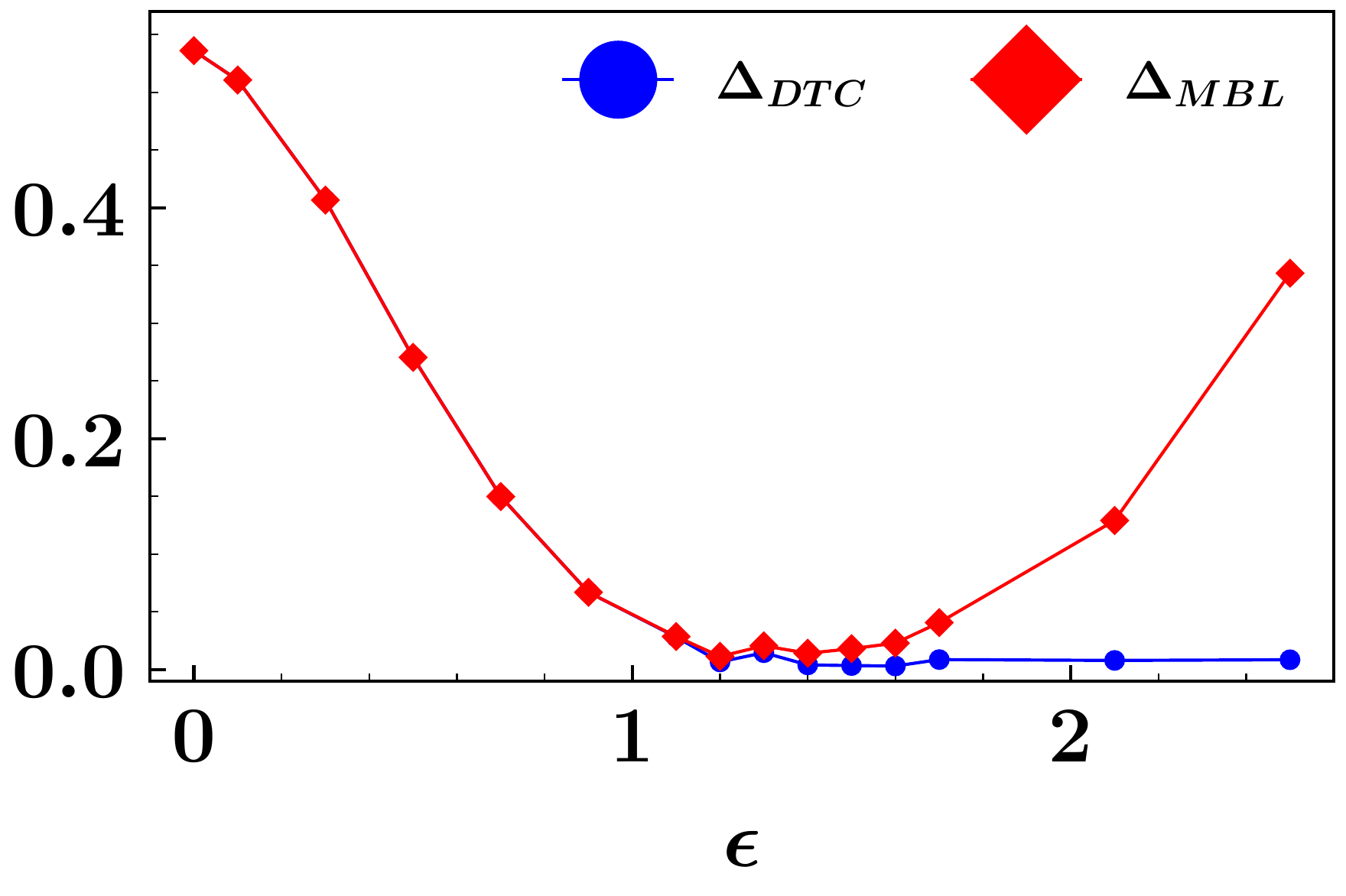}
        \put(-210.5, 120.0){\scalebox{1.5}{\textbf{(f)}}}
    \end{minipage}

    \caption{\justifying (a) Entanglement entropy $S_{ent}(nT)$ as a function of time for rotation-angle errors $\epsilon \in [0.0,2.5]$, for chain length $N=12$. The dotted horizontal line is the maximum entropy value allowed by the chosen bipartition for the finite chain.
    (b) uniform-magnetization QFI per spin $F_Q(nT)$, (c) Connected nearest-neighbor correlation function $C_{56}(nT)$ and (d) connected next-nearest-neighbor correlation function $C_{46}(nT)$ for various rotation-angle errors $\epsilon$ in the three different regimes: $\epsilon=0.3$ (red downward triangles), $\epsilon=1.2$ (blue upward triangles), and $\epsilon=2.5$ (green circles), with chain length $N=11$.
    (e) Operational lower bound $\mathcal E_{\mathrm{ex}}(nT)$ on the restricted-control ergotropy, obtained from the return protocol, for $\epsilon=0.3$ (red), $\epsilon=1.2$ (blue), and $\epsilon=2.5$ (green) for chain length $N=12$. 
    (f) $\Delta_{DTC}$ (blue circles) and $\Delta_{MBL}$ (red squares) as functions of rotation-angle error $\epsilon$, for chain length $N=8$.
    Results are obtained by averaging over $5 \cdot 10^3$ independent disorder realizations, interaction strength $J_z/J_{\perp}=1$, disorder strength $W/J_{\perp}=10$ and driving period $T=2/J_{\perp}$.}
\label{fig:5}
\end{figure*}
The main results for these quantities are reported in Fig.\,\ref{fig:5}. Their dependence on the rotation-angle error reveals three distinct dynamical regimes. For small errors, $\epsilon  \lesssim0.7$, the dynamics clearly exhibits DTC behavior. For errors $0.7 \lesssim \epsilon \lesssim 1.2$, an intermediate regime appears, which we will refer to as Anderson-like. Finally, for large errors, $\epsilon \gtrsim 1.2$, the dynamics are consistent with that of a Floquet-MBL system, namely a driven MBL system in which no DTC phase is established.

Let us first focus on the entanglement entropy. In the low-error DTC regime, it exhibits a logarithmic growth in time. As the rotation-angle error increases, in the interval $0.7 \lesssim \epsilon \lesssim 1.2$, the entropy initially increases but subsequently saturates to an approximately stationary value. Appendix\,\ref{app:N} shows that this behavior is not due  to finite-size effects. For even larger values of $\epsilon$, the entanglement entropy reaches lower values and again exhibits slow localized dynamics.

In our view, the saturation observed in the intermediate regime is reminiscent of Anderson localization. Unlike conventional Anderson localization, which arises from the absence of interactions, this regime appears to be induced by the interplay between longitudinal interactions, disorder, and the effect of the periodic kicks. Their competition effectively suppresses the propagation of correlations, resulting in an Anderson-like saturation of the entanglement entropy. 

The same result holds for the uniform-magnetization QFI shown in Fig.\,\ref{fig:5} (b). Its temporal evolution with respect to the rotation-angle error is qualitatively similar to that of the entanglement entropy: the QFI grows in the low- and high-error localized regimes, whereas it saturates in the intermediate regime.  The saturation of the QFI indicates that the collective fluctuations of the magnetization become approximately stationary, providing further evidence for the suppression of long-range dynamical correlations in the intermediate regime.

Meanwhile, panel (c) to (f) underline the analogy between the intermediate regime in the rotation angle $\epsilon\approx 1.2$ and the Anderson localized phase in absence of external drive ($\hat{U}_F(T)=e^{- i \hat{H}_{XXZ} T}$). They show the three different regimes of rotation angle error for $J_z/J_{\perp}=1.0$ and then also the case Anderson $J_z/J_{\perp}=0.0$ case. In all the quantities, two-point correlation functions $G_2$, entanglement entropy $S_{ent}$ and spatial correlation functions for nearest and next-nearest neighbourhood, the Anderson-undriven case manifests a stationary time behaviour which is analogous to the corrispective intermediate regime trend.

Figures \ref{fig:5} (c) and \ref{fig:5} (d) show the connected correlation functions in time of the central site spin ($i=6$) with the nearest- and next-nearest-neighbor spins ($C_{56}(t)$ and $C_{46}(t)$, respectively). These quantities describe correlations between fluctuations of the longitudinal spin components. A positive value of $C_{ij}$ indicates that the longitudinal fluctuations of the two spins tend to have the same sign, whereas a negative value indicates that they tend to have opposite signs. The opposite signs of $C_{56}$ and $C_{46}$ can be understood in terms of the staggered structure inherited from the initial Néel configuration. Sites $i=5$ and $i=6$ belong to opposite sublattices, whereas sites $i=4$ and $i=6$ belong to the same sublattice. The signs of the connected correlations therefore reflect the alternating pattern of the correlated longitudinal fluctuations. In the low- and high-error localized regimes, $C_{56}(t)$ and $C_{46}(t)$ remain appreciably different from zero. In the intermediate regime, both correlations become strongly suppressed and approach zero. This means that the longitudinal fluctuations of these selected pairs become approximately uncorrelated.

Nevertheless, the simultaneous suppression of the selected short-range longitudinal correlations, the saturation of the uniform QFI, and the saturation of the entanglement entropy provide mutually consistent evidence for an effectively weakly correlated, Anderson-like dynamical regime.

\begin{figure*}[]
  \centering

    \begin{minipage}{0.48\textwidth}
        \centering
        \includegraphics[width=\linewidth]{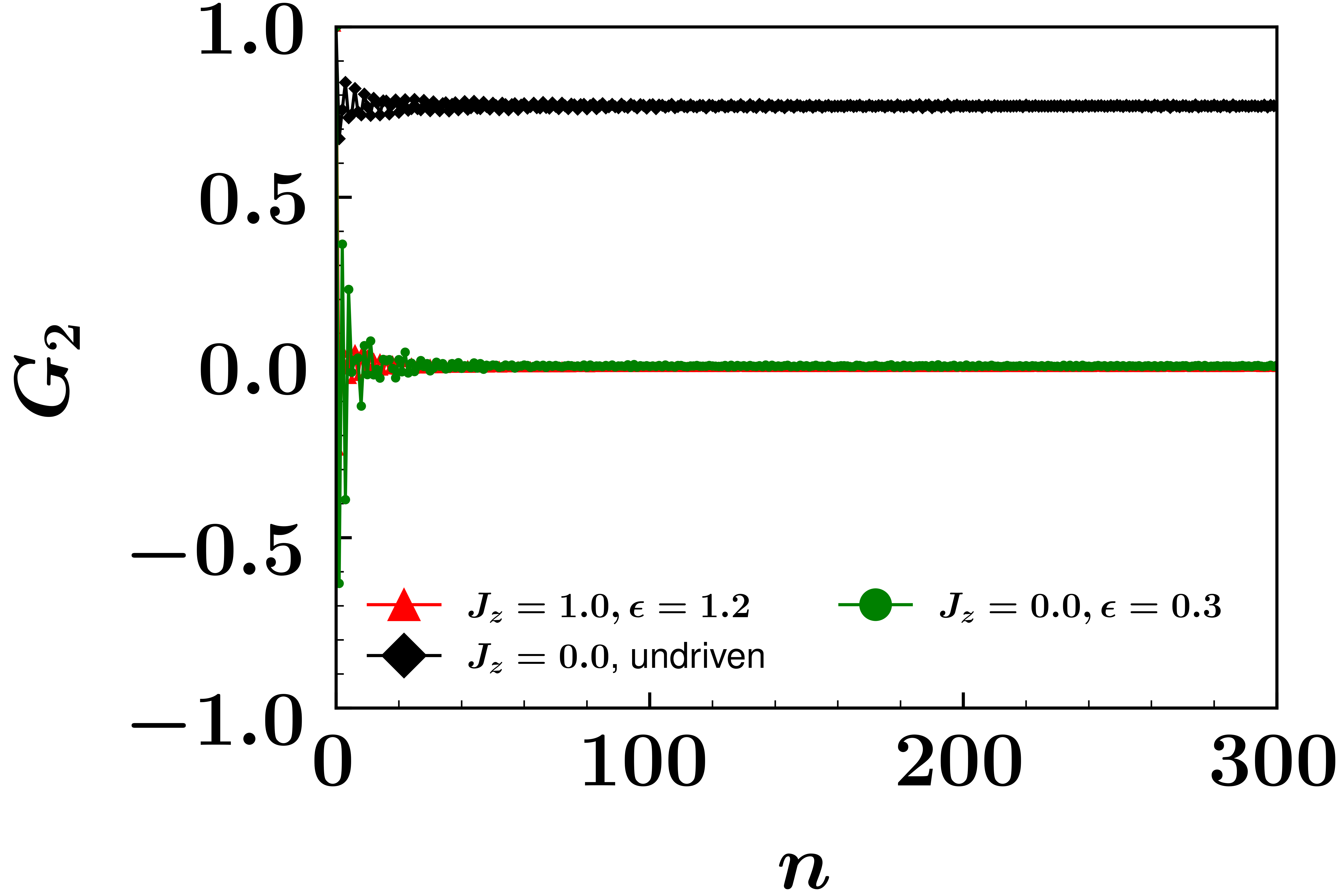}
         \put(-220.5, 120.0){\scalebox{1.5}{\textbf{(a)}}} 
    \end{minipage}
    \hfill
    \begin{minipage}{0.48\textwidth}
        \centering
        \includegraphics[width=\linewidth]{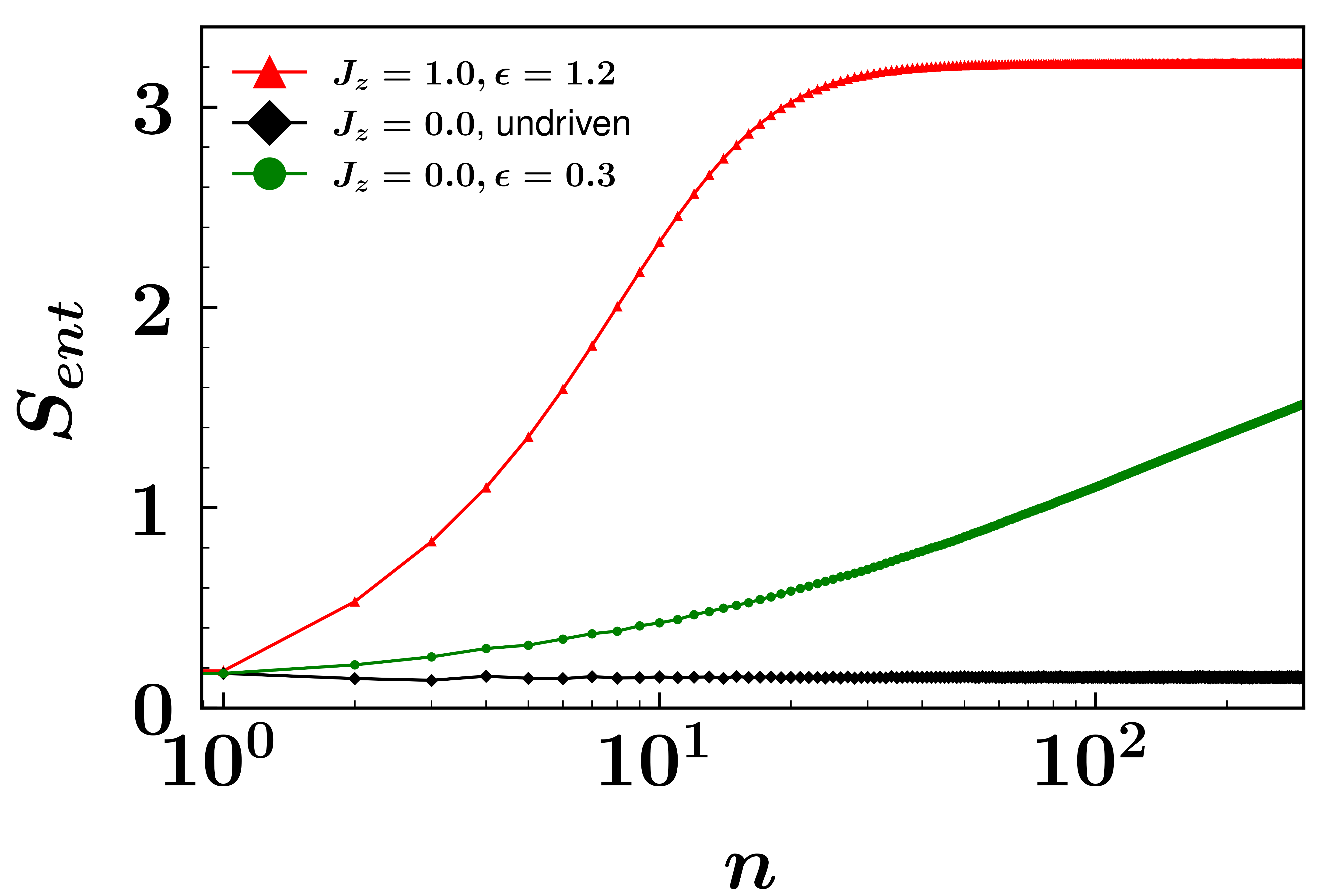}
       \put(-210.5, 120.0){\scalebox{1.5}{\textbf{(b)}}}
    \end{minipage}

    \vspace{0.0cm}

    \begin{minipage}{0.48\textwidth}
        \centering
        \includegraphics[width=\linewidth]{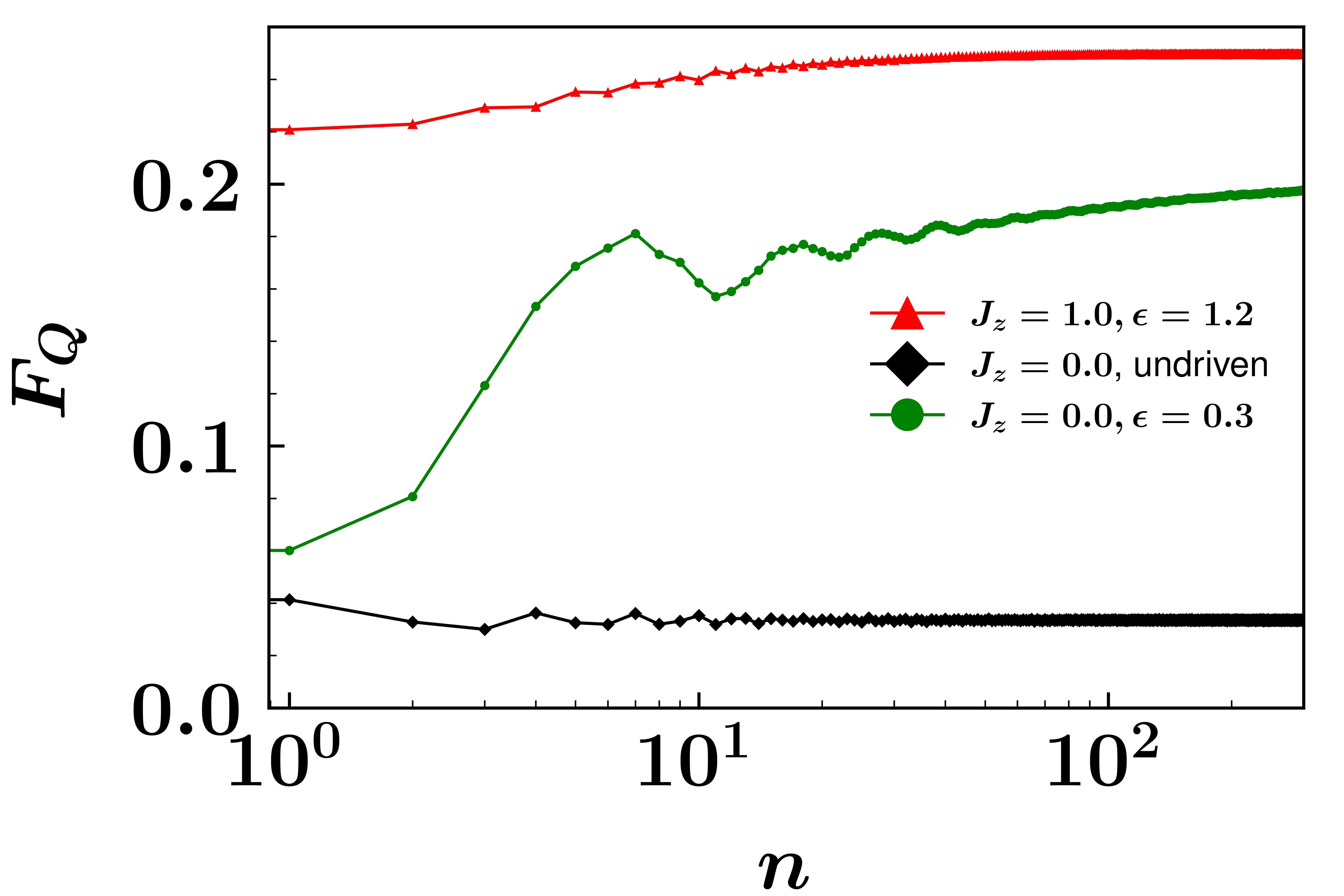}
        \put(-220.5, 120.0){\scalebox{1.5}{\textbf{(c)}}}
    \end{minipage}
    \hfill
    \begin{minipage}{0.48\textwidth}
        \centering
        \includegraphics[width=\linewidth]{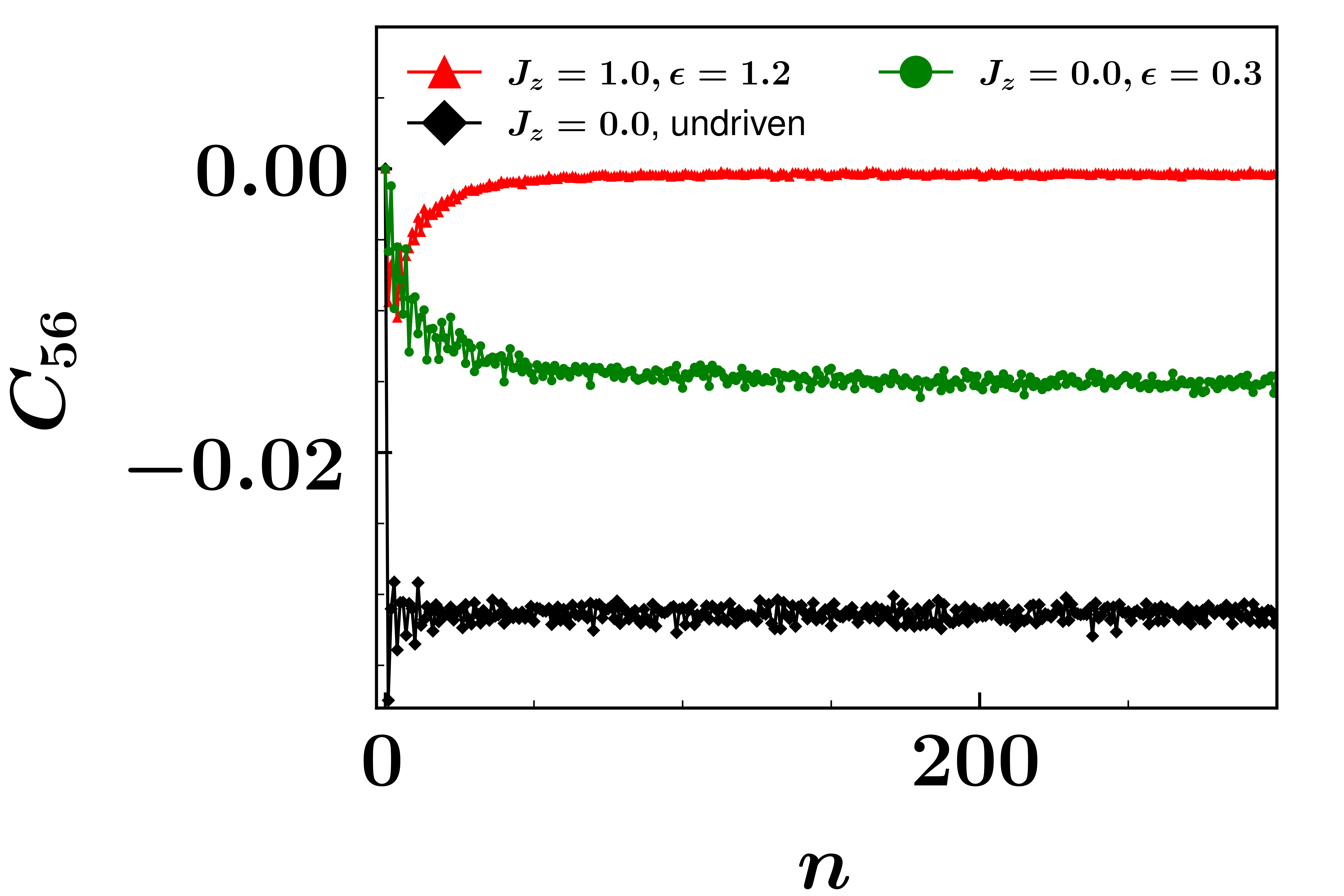}
         \put(-210.5, 120.0){\scalebox{1.5}{\textbf{(d)}}}
    \end{minipage}
\caption{\justifying (a) Two-time imbalance correlation function $G_2(0,nT)$ (b) Entanglement entropy $S_{ent}(nT)$, (c) quantum Fisher information $F_Q(nT)$ and (d) Connected nearest-neighbor correlation function $C_{56}(nT)$ as a function of the number $n$ of Floquet cycles, in three distinct cases: $J_z/J_{\perp}=1$ with $\epsilon=1.2$ (red triangles),  $J_z/J_{\perp}=0$ in the absence of drive (black diamonds) and  $J_z/J_{\perp}=0$ with $\epsilon=1.2$ (green circles).  Results are obtained by averaging over $5 \cdot 10^3$ independent disorder realizations with disorder strength $W/J_{\perp}=10$ and driving period $T=2/J_{\perp}$ for a chain length $N=11$.}
\label{fig:6}
\end{figure*}

This interpretation of the underlying physics in the intermediate-error regime is further supported by the results shown in Fig.\,\ref{fig:6}, where we report the time evolution of the imbalance correlation function $G_2(0,nT)$ (panel a), the entanglement entropy $S_{\mathrm{ent}}(nT)$ (panel b), quantum Fisher information $F_Q(nT)$ (panel c) and the connected nearest-neighbor correlation function $C_{56}(nT)$ (panel d) for several cases. Specifically, we consider the interacting case with $J_z/J_{\perp}=1$ in the intermediate-error regime, the case with $\epsilon=1.2$, and the Anderson-localized regime with $J_z=0$, both in the presence and absence of the external drive.
The three panels reveal that the undriven Anderson-localized case exhibits a remarkably similar behavior to the driven $J_z/J_{\perp}=1$ system in the intermediate-error regime. In both cases, the observables become approximately stationary at long times, indicating a strong connection between the two dynamical regimes.

In the absence of interactions ($J_z=0$) and driving, the imbalance correlation function remains at relatively high values, as the system evolves only weakly away from its initial state. Consistently, the entanglement entropy is small and it remains essentially constant, indicating a lack of significant entanglement growth. The quantum Fisher information follows a behaviour similar to the entanglement entropy.  
In contrast, in the driven Anderson-localized case, although the imbalance correlation function displays small oscillations around zero, similarly to the driven $J_z/J_{\perp}=1$ case, the nearest-neighbor correlations exhibit a slight decrease over time, while the entanglement entropy and quantum Fisher information increase slightly. These trends can be attributed to the action of the external drive. 
These results further suggest that the dynamics observed in the intermediate-error regime share relevant features with those of an Anderson-localized system, while the external drive induces a weak but non-negligible development of correlations.

A behaviour similar to those of the QFI and entanglement entropy is observed for the operational lower bound on the restricted-control ergotropy in Fig.\,\ref{fig:5} (e). As the rotation-angle error increases, $\mathcal E_{\mathrm{ex}}$ initially grows, reaching the largest values in the intermediate regime, before decreasing in the large-error Floquet-localized regime. The presence of oscillations that are not perfectly coherent in the energy signal within the DTC phase should not be regarded as anomalous. DTC order is identified through local observables that transform nontrivially under discrete time translations and consequently exhibit a rigid subharmonic response. The imbalance is such an observable because an ideal global spin flip reverses its sign. By contrast, the different terms of the Hamiltonian transform differently under the global spin rotation and the energy difference is therefore not required to oscillate with period $2T$. More details about the analogy between Anderson localization and rotation error intermediate regime can be found in Appendix.\,\ref{app:J_z}.

Finally, Fig.\,\ref{fig:5} (f) shows a more detailed analysis of the three distinct dynamical regimes, DTC, Anderson-like and FLoquet-MBL, using the signatures defined in Eq.\,\eqref{DELTA_eq}. We note that for small $\epsilon$ both the values of $\Delta_{MBL}$ and $\Delta_{DTC}$ are nonzero, indicating that the system retains memory of its initial configuration while displaying a stable alternating response. This behavior is consistent with a localization-stabilized DTC regime. In the intermediate regime, both indicators are strongly suppressed. In the third regime, $\Delta_{MBL}$ becomes nonzero while $\Delta_{DTC}$ remains close to zero. This indicates localized Floquet dynamics without a stable time-crystalline response.  These results complement previous behaviors discussed in the literature \cite{switzer2026realization}. 


More details about the robustness of DTC phase can be found in Appendices.

\section{Conclusions}

Using matrix-product-state simulations, we have investigated time-crystalline dynamics in a disordered Heisenberg chain whose undriven parameters are chosen within the many-body-localized regime and which is subjected to a periodic sequence of global spin rotations generated by delta kicks. For an ideal pulse, the rotation angle is $\pi$ about the $x$ axis, reversing the $z$ component of each spin.

Our results provide evidence that robust time-crystalline behavior can persist in a strongly disordered one-dimensional Heisenberg chain, including at the isotropic point, under a minimal Floquet protocol consisting of a single global spin rotation per driving period. This behavior is obtained without additional interaction-engineering pulses, strong field gradients, or nonstandard lattice geometries. We have further investigated the robustness of the time-crystalline response against rotation-angle errors using observables from quantum information, many-body localization, and quantum thermodynamics.

For small rotation-angle errors, the dynamics retains localization signatures together with a robust period-doubled response. In particular, the imbalance correlation function exhibits oscillations with a period twice that of the external drive. As the rotation angle deviates from the ideal $\pi$ pulse, the spectral weight of the subharmonic peak progressively decreases. For large rotation-angle errors, the dynamics again displays signatures compatible with Floquet localization, including slow entanglement growth, while there is no longer evidence of time-crystalline behavior.

One of our main results is the observation of an intermediate dynamical regime as a function of the rotation-angle error. In this region, the imbalance correlation function becomes approximately stationary, indicating that the period-doubled response has disappeared, while the dynamics differs qualitatively from the localized regimes observed at smaller and larger errors. Selected connected longitudinal spin correlations are strongly suppressed, while the quantum Fisher information and entanglement entropy approach stationary values. Taken together, these signatures are consistent with an effectively weakly correlated intermediate dynamical regime whose behavior is reminiscent of Anderson localization. 

Building on these results, an important direction for future work is to extend this framework to open systems, where MPS-based numerical methods \cite{grazia3,grazia1,grazia2} could be used to investigate the stability of time-crystalline dynamics against dissipation and decoherence. Such studies would provide further insight into the interplay between periodic driving, many-body interactions, disorder, and environmental effects \cite{kongkhambut2022observation,wu2024dissipative,wu2026dissipative,russo2025quantum,khasseh2026semiclassical}. Another natural direction is to investigate whether the intermediate weakly correlated regime identified here persists in higher-dimensional Heisenberg-type systems, including settings related to the two-dimensional geometry studied in Ref.~\cite{switzer2026realization}. This would help clarify the roles of dimensionality, disorder, interactions, driving, and dissipation in stabilizing or suppressing time-crystalline dynamics.

\section*{Acknowledgements}
F.F., G.D.B. and C.A.P. acknowledge funding from IQARO (Spin-orbitronic Quantum Bits in Reconfigurable 2DOxides) project of the European Union’s Horizon Europe research and innovation programme under grant agreement n. 101115190. 
C.A.P. and G.D.F. acknowledge financial support from PNRR MUR Project No. PE0000023-NQSTI. F.F. also acknowledges the CINECA award under the ISCRA initiative (project HP10C7PORQ), for the availability of high performance computing resources and support.

\appendix

\section*{Appendix}
We add some details relevant to complete the analysis of the DTC features of the Heisenberg chain. We study the system by varying the chain length $N$, the disorder strength $W$, the period $T$ of the drive, and the interaction strength $J_z$. Finally, we recall the Jordan-Wigner transformations.  

\section{Variation of the chain length $N$}\label{app:N}
In this section, we focus on differences in the time behavior of the two-time imbalance correlation function and entanglement entropy for different chain lengths $N$.
\begin{figure}[h]
\centering
  \includegraphics[width=0.5\columnwidth]{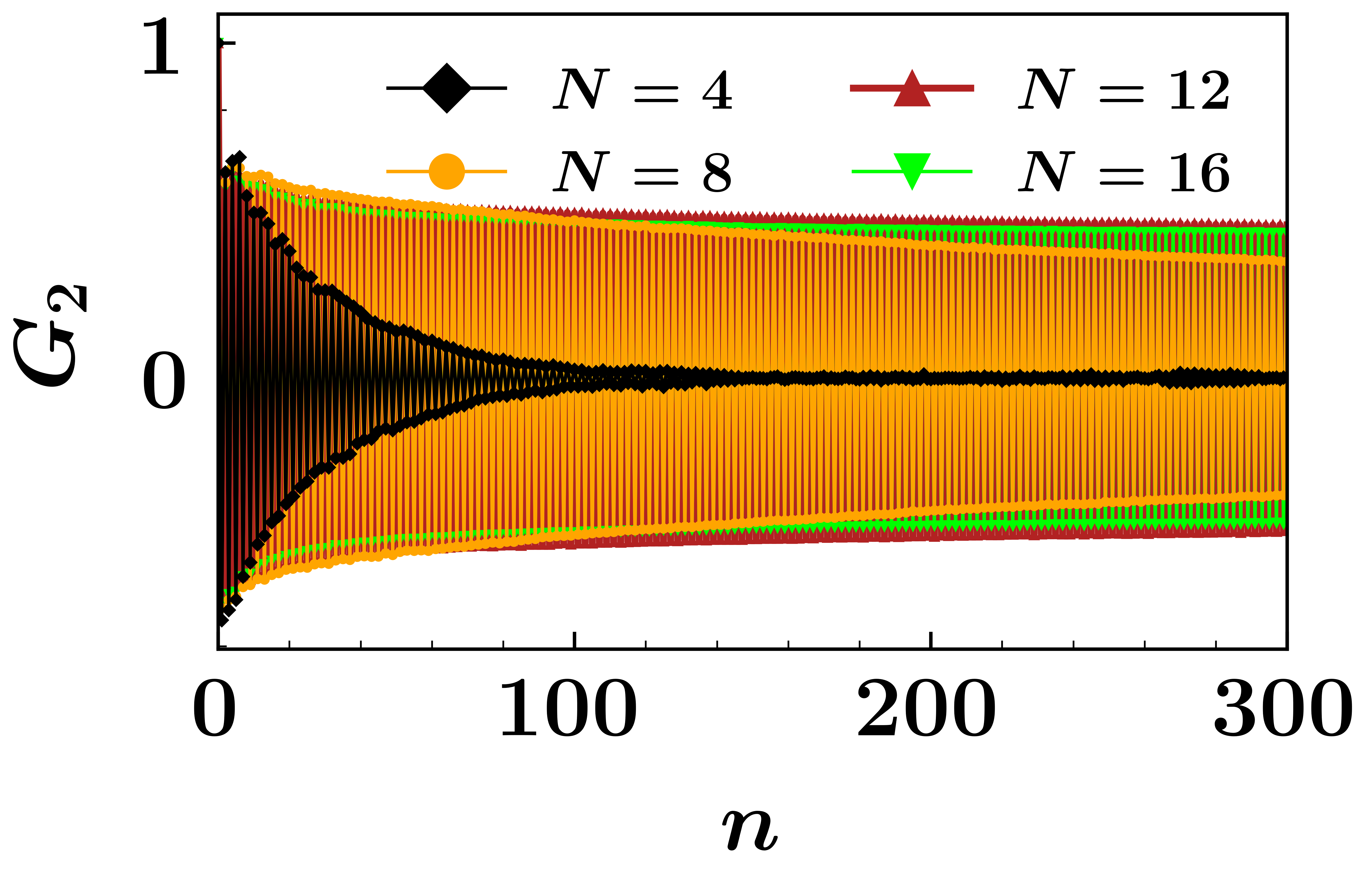}
\caption{\justifying Two-time correlation function $G_2(0,n T)$ as a function of the number $n$ of Floquet cycles for different chain lengths: $N=4$ (black), $N=8$ (orange), $N=12$ (firebrick), and $N=16$ (lime). Results are obtained for disorder strength $W/J_{\perp}=10$, longitudinal interaction $J_z/J_{\perp}=1.0$, and period $T=2/J_{\perp}$, and by averaging over $5 \cdot 10^3$ independent disorder realizations, except $N=16$, for which the average is performed over $3 \cdot 10^3$ realizations.}
  \label{fig:app:1}
\end{figure}

In Fig.\,\ref{fig:app:1}, we show the two-time imbalance correlation function $G_2$ for various chain lengths $N$. For the smallest system, $N=4$ (the case analysed in \cite{switzer2026realization}), the signal rapidly decays to zero, indicating that the DTC phase is highly unstable. As $N$ increases, the phase becomes progressively more stable. 
Indeed, for $N\ge 12$, the signal appears to have reached a very stable behaviour for the chosen set of parameters, with the remaining fluctuations attributable to statistical errors arising from averaging over different disorder realizations.

Based on the data reported in Fig.\,\ref{fig:app:1}, we conducted an analysis on the amplitude decay of the $G_2(0,nT)$ signal. For each chain length $N$, we characterize the temporal decay by fitting the data to an exponential-law function $G_2(n)=Ae^{-n/\tau}+C$, primary interested in the decay times $\tau$.
We found that the best parameters that fit the data are $\tau=28$ for $N=4$, $\tau=501$ for $N=8$, $\tau=1252$ for $N=12$, and, finally, $\tau=1142$ for $N=16$. These results indicate that the signal amplitude exhibits a persistent behaviour as the chain length increases and, specifically, for $N\ge12$ it goes stabilizing. 
The extracted decay times are substantially larger than the simulated time window  ($n=300$) and should therefore be interpreted only as lower-scale indicators of the long-lived character of the response.

\begin{figure}[h]
\centering
  \includegraphics[width=0.5\columnwidth]{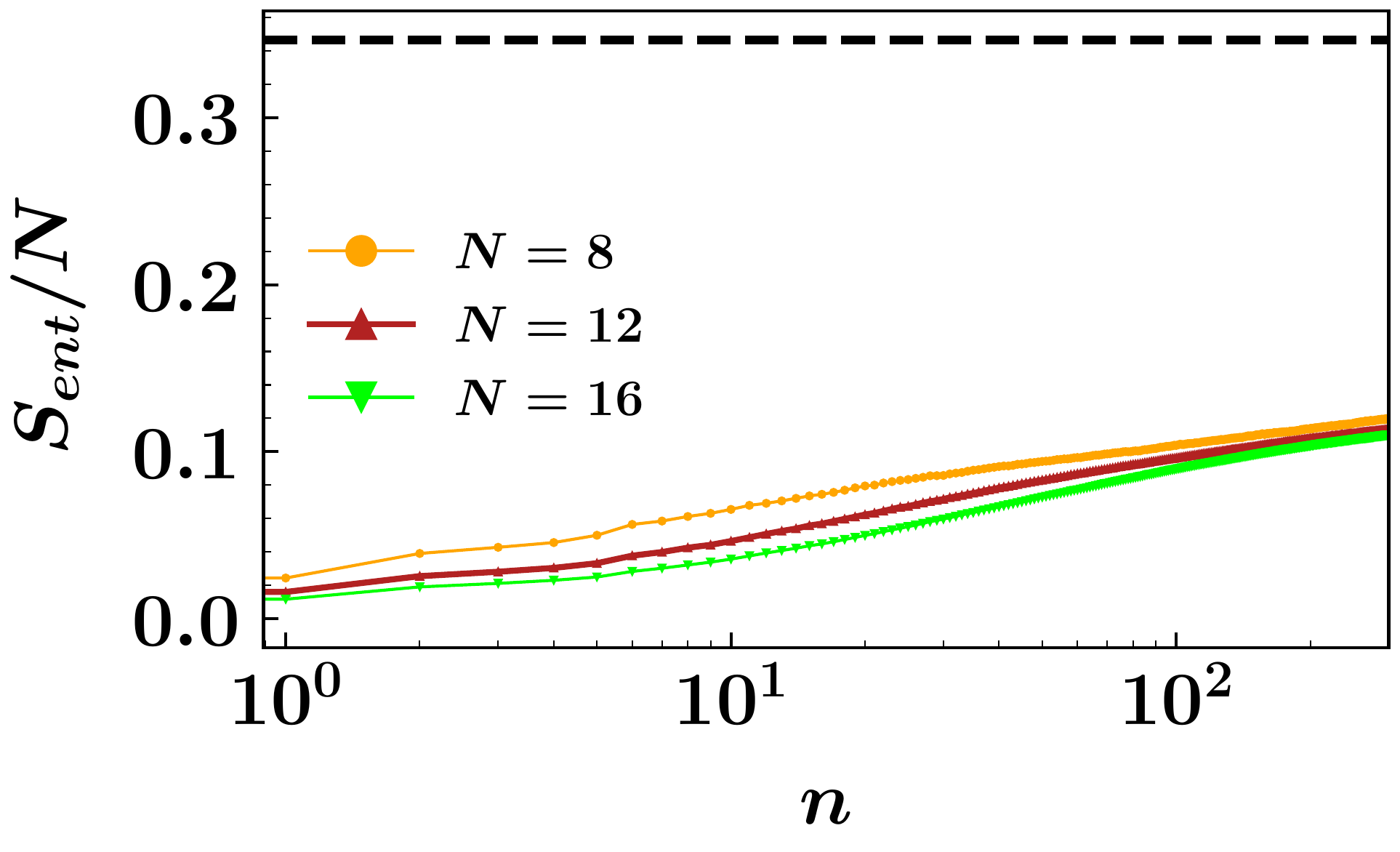}
\caption{\justifying Entanglement entropy per site 
as a function of the number $n$ of Floquet cycles for different chain lengths: $N=8$ (orange circles), $N=12$ (firebrick upward triangles), and $N=16$ (lime downward triangles). The dashed line are the maximum values of the entropy per site allowed for each length $N$ (all lines overlap). Results are obtained for disorder strength $W/J_{\perp}=10$, longitudinal interaction $J_z/J_{\perp}=1.0$, and period $T=2/J_{\perp}$, and by averaging over $5 \cdot 10^3$ independent disorder realizations, except for $N=16$, for which the average is performed over $3 \cdot 10^3$ realizations.}
  \label{fig:app:3}
\end{figure}

Furthermore, Fig.\,\ref{fig:app:3} shows the time evolution of the entanglement entropy for three different chain lengths $N$, together with the corresponding maximum value $S^{max}_{ent}=\frac{N}{2}\log(2)$ (dotted lines) allowed for each finite system size. As can be seen, the maximum value reached by the entanglement entropy during the dynamics remains well below $S^{max}$ in all cases. This indicates that the observed saturation is not simply caused by the finite-size upper bound on the entropy.

\section{Variation of the disorder strength $W$}\label{app:W}
\begin{figure}[h]
\centering
  \includegraphics[width=0.5\columnwidth]{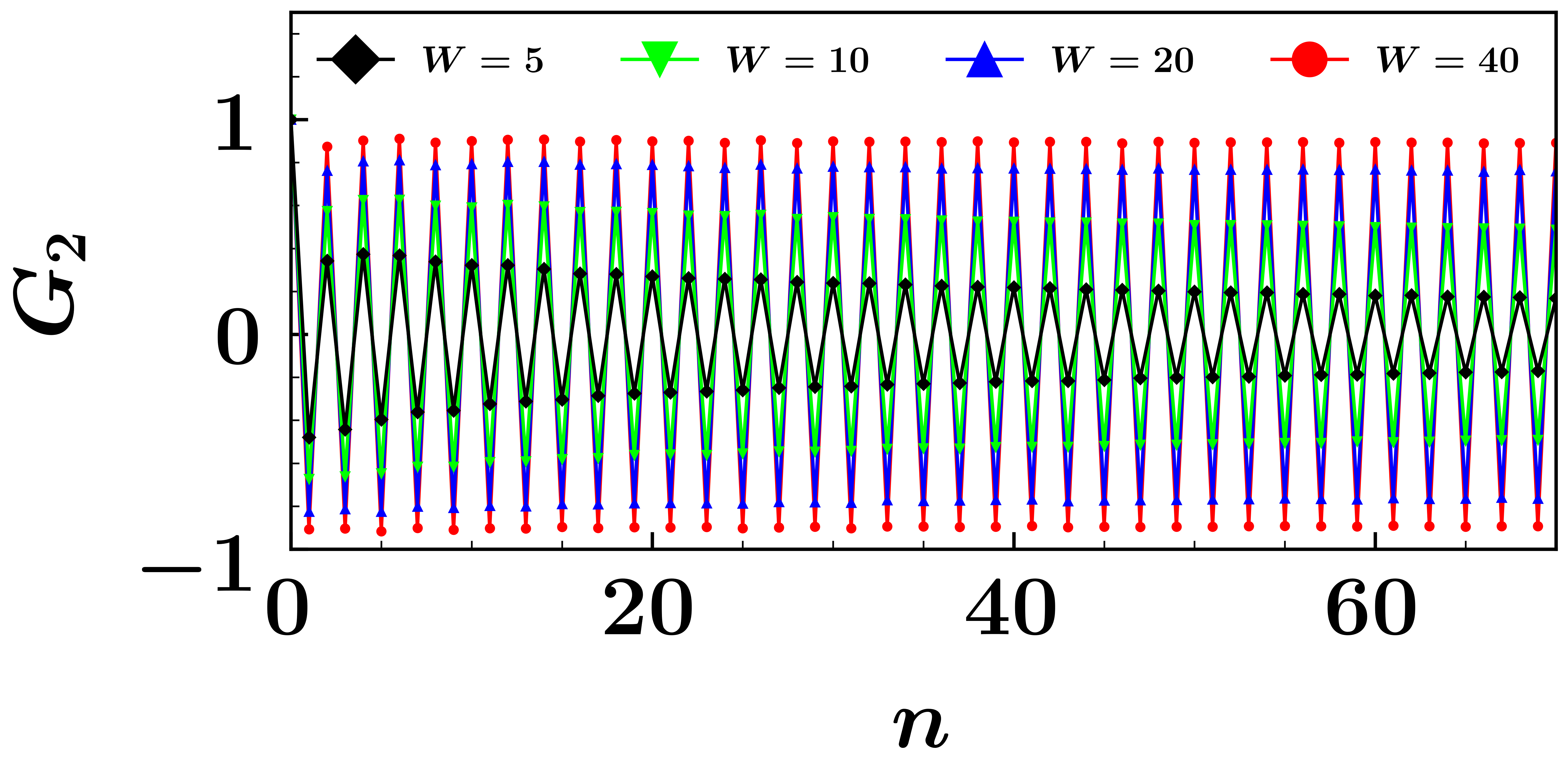}
\caption{\justifying Two-time correlation function $G_2(0,n T)$ as a function of the number $n$ of Floquet cycles for different values of disorder strengths: $W/J_{\perp}=5$ (black diamonds), $W/J_{\perp}=10$ (lime downward triangles), $W/J_{\perp}=20$ (blue upward triangles), and $W/J_{\perp}=40$ (red circles). 
Results are obtained for driving period $T=2/J_{\perp}$, longitudinal interaction $J_z/J_{\perp}=1.0$, and chain length $N=8$, and by averaging over $5 \cdot 10^3$ independent disorder realizations.}
  \label{fig:app:4}
\end{figure}
In this section, we analyse differences in the dynamical behavior of the two-time imbalance correlation function for different disorder strengths $W /J_{\perp}$. Specifically, we observe in Fig.\,\ref{fig:app:3} that the larger $W$ is, the larger the amplitude of the period-doubled oscillations. This can be understood physically as follows. The stronger the disorder, the more strongly the dynamics of the underlying MBL phase are enhanced, meaning that the system remains closer to its initial state throughout the time evolution. As a result, the z-component of each spin is less likely to change due to the intrinsic dynamics, allowing the drive to flip it with higher fidelity.

\section{Variation of the period T}\label{app:T}

In this section, we analyse differences in the dynamical behavior of the two-time imbalance correlation function and the entanglement entropy for different values of the dimensionless driving period $T J_{\perp}$. Specifically, we show that, for the range of parameters $\{J_z, W, N\}$ analysed in the main text, for the parameters and time window considered, values $T \gtrsim J^{-1}_{\perp}$ are required to retain  localization dynamics. 
\begin{figure}[h]
\centering
  \begin{overpic}[width=0.5\columnwidth]{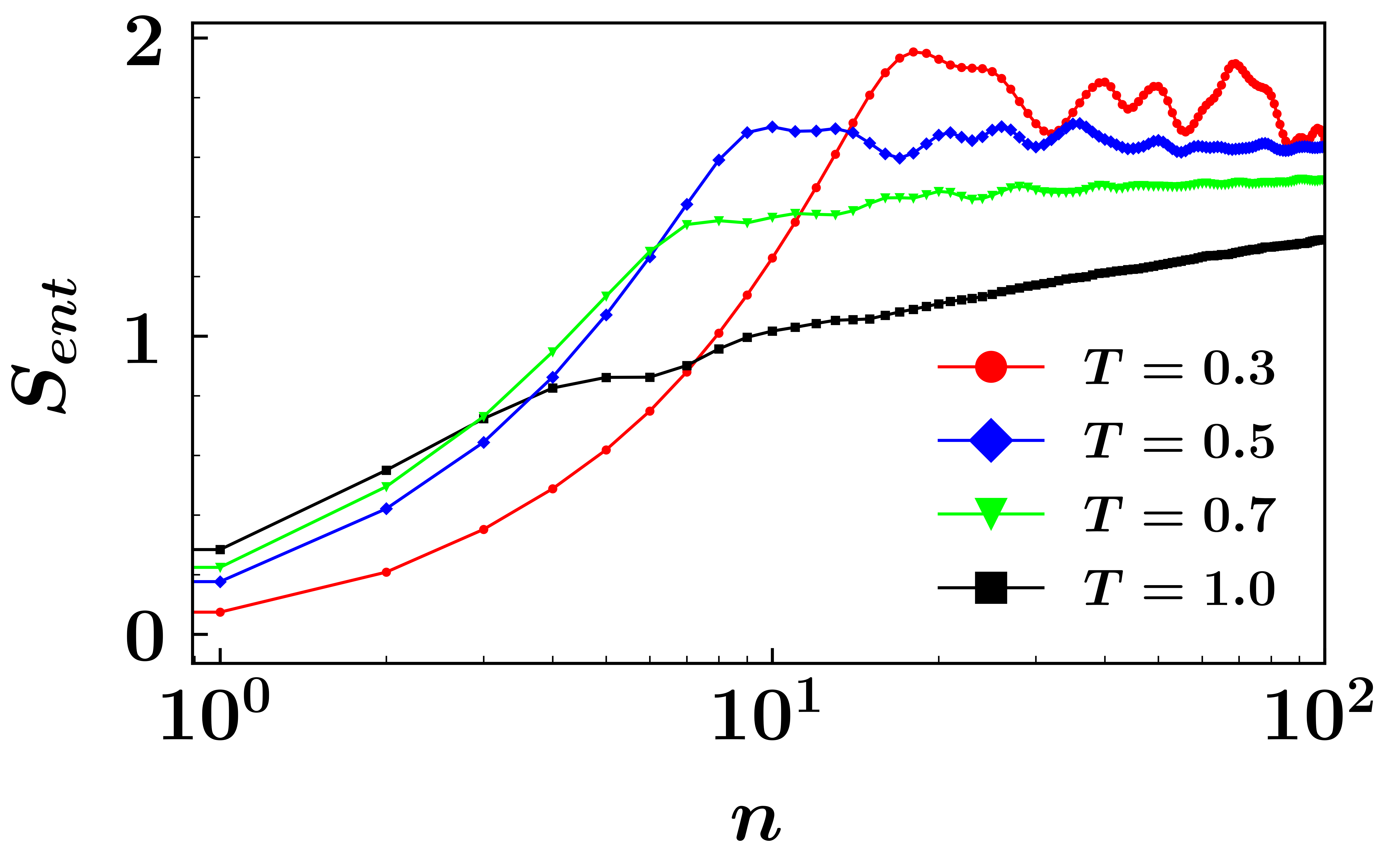}
    \put(-3.2, 57.0){\scalebox{1.5}{\textbf{(a)}}} 
  \end{overpic}
  \hspace{1.7cm}  
  \begin{overpic}[width=0.5\columnwidth]{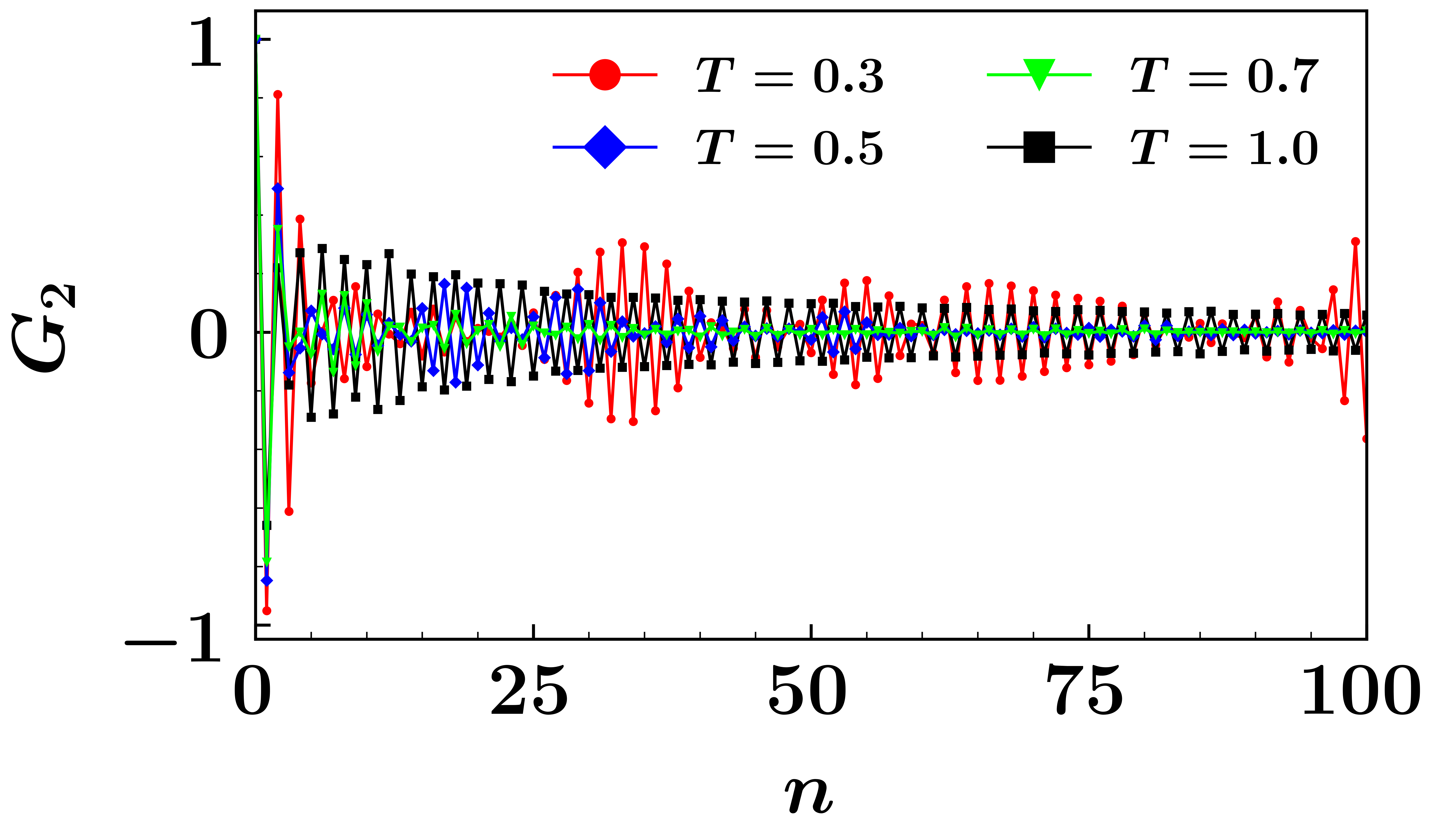}
   \put(-3.2, 57.0){\scalebox{1.5}{\textbf{(b)}}}  
  \end{overpic}
   \caption{\justifying (a) Entanglement entropy $S_{ent}$ and (b) two-time correlation function $G_2(0,nT)$ as a function of the number $n$ of Floquet cycles for different values of the driving period: $T=0.3/J_{\perp}$ (red circles), $T=0.5/J_{\perp}$ (blue diamonds), $T=0.7/J_{\perp}$ (lime downward triangles), and $T=1.0/J_{\perp}$ (black squares). Results are obtained by averaging over $5 \cdot 10^3$ independent disorder realizations, with disorder strength $W/J_{\perp}=5$, longitudinal interaction $J_z/J_{\perp}=1.0$, and chain length $N=8$.}
\label{fig:app:5}
\end{figure}

We now analyse the entanglement entropy time behavior for different values of $T$. It is known that MBL systems show a logarithmic increase in entanglement over time, while, in the ergodic phase, entanglement increases ballistically until it reaches the finite-size maximum value (oscillating around it). In panel (a) of Fig.\,\ref{fig:app:5}, we see that, once the driving period falls below $T=J^{-1}_{\perp}$, the entanglement entropy grows at an increasingly faster rate. For $T \gtrsim J^{-1}_{\perp}$ the logarithmic behavior is restored. In panels (b) and (c) of Fig.\,\ref{fig:app:5} we observe the two-time correlation function $G_2$ and its frequency spectrum, respectively. For $T< J^{-1}_{\perp}$, the coherence of period-doubled oscillations disappears, so the system can no longer be considered to be in DTC phase.

\section{Variation of interaction strength $J_z$}\label{app:J_z}

In this section, we characterize differences in the dynamical behavior of the two-time imbalance correlation function for different values of the longitudinal interaction strength $J_z/J_{\perp}$.

\begin{figure*}[]
    \centering

    \begin{minipage}{0.48\textwidth}
        \centering
        \includegraphics[width=\linewidth]{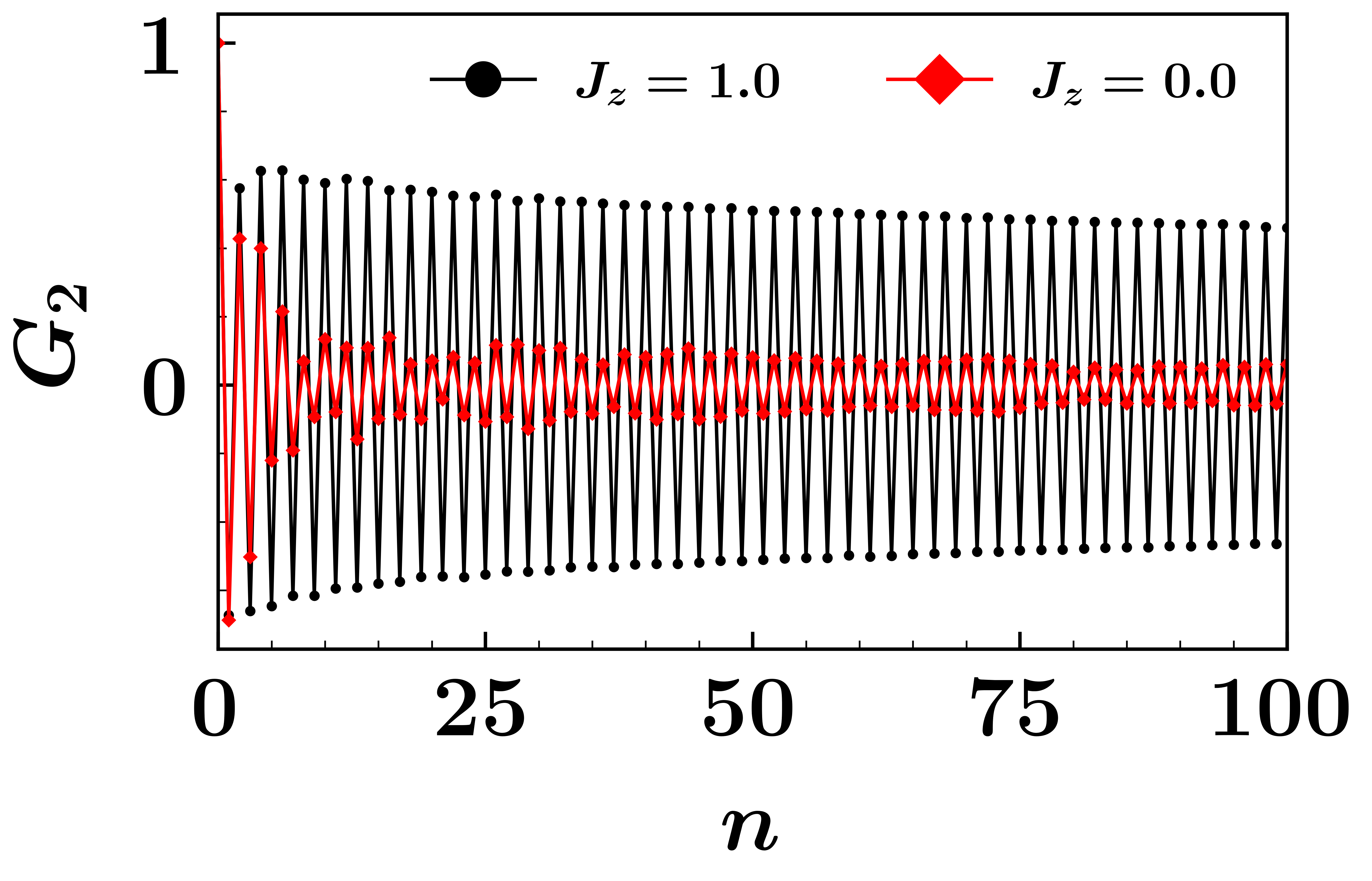}
         \put(-220.5, 120.0){\scalebox{1.5}{\textbf{(a)}}} 
    \end{minipage}
    \hfill
    \begin{minipage}{0.48\textwidth}
        \centering
        \includegraphics[width=\linewidth]{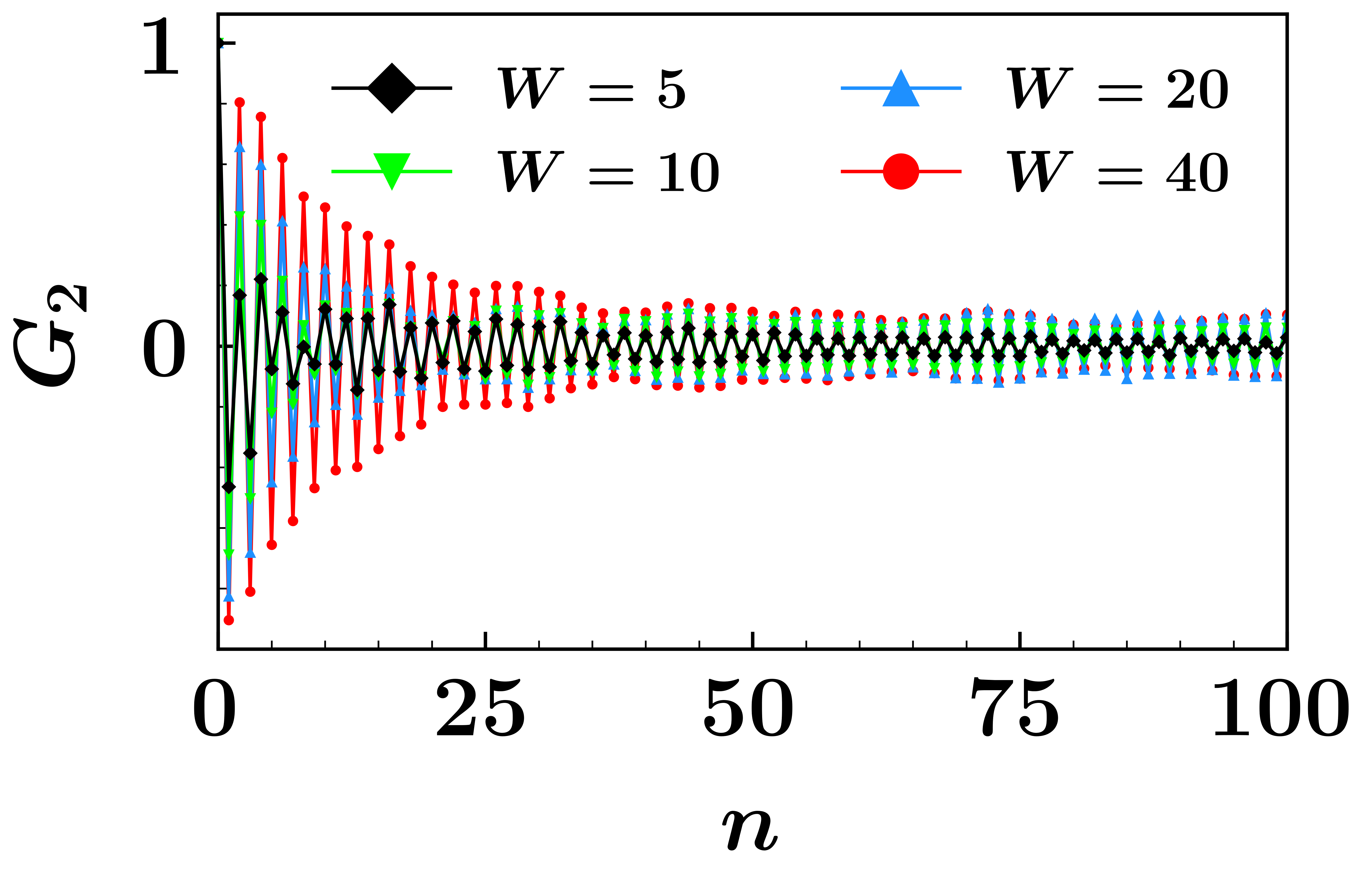}
       \put(-210.5, 120.0){\scalebox{1.5}{\textbf{(b)}}}
    \end{minipage}
    \caption{\justifying (a) Two-time correlation function $G_2(0,nT)$ as a function of the number $n$ of Floquet cycles in the Anderson-localized regime, $J_z/J_{\perp}=0.0$ (black circles), and in the MBL regime $J_z/J_{\perp}=1.0$ (red diamonds), for fixed $W/J_{\perp}=10.0$. (b) $G_2(0,n T)$ as a function of the number $n$ of Floquet cycles for different values of the disorder strength: $W/J_{\perp}=5.0$ (black diamonds), $W/J_{\perp}=10.0$ (green downward triangles), $W/J_{\perp}=20.0$ (blue upward triangles), and $W/J_{\perp}=40.0$ (red circles), for fixed $J_z/J_{\perp}=0.0$ in the Anderson-localized regime. All results are obtained for chain length $N=8$ by averaging over $5 \cdot 10^3$ independent disorder realizations. }
\label{fig:app:6}
\end{figure*}

We show that, when the system is Anderson localized ($J_z=0$) and MBL cannot occur, the system does not exhibit any DTC behavior.
In the panel (a) of Fig.\,\ref{fig:app:6}, we report two different $G_2(0,nT)$ curves. For $J_z=0$, the oscillations are not coherent, while, for $J_z/J_{\perp}=1$, the subharmonic response is preserved for long times. The panel (b) of Fig.\,\ref{fig:app:6} also shows that this behavior persists over the range of disorder strengths considered.

\section{Jordan-Wigner transformations}\label{app:Jordan-Wigner}
The chain analysed is a one-dimensional anisotropic disordered Heisenberg model with nearest-neighbor interactions, open boundary conditions and randomly sampled external on-site fields. It is useful to underline that the physics of this spin system is the same of a one-dimensional model of interacting disordered spinless fermions with open boundary conditions:
\begin{equation}\label{eq:hsyst}
\hat{H}= -t \sum\limits^{{N-1}}_{i=1}
    \bigg\{\hat{c}^{\dag}_i \hat{c}_{i+1} + \hat{c}^{\dag}_{i+1} \hat{c}_i \bigg\} +  {V} \sum\limits^{N-1}_{i=1} \hat{n}_{i}\hspace{0.5mm}\hat{n}_{i+1}+\sum\limits^{N}_{i=1}H_i \hspace{1mm}\hat{n}_{i},
\end{equation}
 where, referring to Eq.\,\eqref{eq:hi} in the main text, $J_{\perp}=-2t$ is the transverse spin-flip interaction energy associated with the hopping $t$, $J_z=V$ is the longitudinal interaction strength, and $h_i=H_i$ are the on-site energy terms associated with the random magnetic fields. The operator $ \hat{c}_i $ ($\hat{c}^{\dag}_i$) annihilates (creates) an electron at the site $i$. The fermionic system is assumed for simplicity to be at half-filling.

The two models are physically equivalent because there is an exact mapping between them. This transformation, known as the Jordan-Wigner transformation, maps “down” and “up” single-spin states onto empty and singly occupied single-fermion states:
\begin{equation}
    \begin{cases}
\ket{\uparrow}\coloneqq\hat{c}^{\dag}\ket{0}\to\hat{S}^{(+)}\coloneqq\hat{c}^{\dag},
    \hspace{2cm}
    \\[2.6ex]
  \ket{\downarrow}\coloneqq\ket{0}\to\hat{S}^{(-)}\coloneqq\hat{c},
\\[2.6ex]
 \hat{S}^{(z)}\coloneqq\hat{n}-\frac{1}{2}=\hat{c}^{\dag}\hat{c}-\frac{1}{2},
 \\[1.7ex]
       \hat{S}^{(x)}=\frac{1}{2}( \hat{S}^{(+)} + \hat{S}^{(-)})=\frac{1}{2}(\hat{c}^{\dag}+\hat{c}),
    \\[1.7ex]
    \hat{S}^{(y)}=\frac{1}{2\mathrm{i}}( \hat{S}^{(+)} - \hat{S}^{(-)})=\frac{1}{2\mathrm{i}}(\hat{c}^{\dag}-\hat{c}).
    \end{cases}
\end{equation}
For multiple spins, spin operators on different sites commute, while fermionic ones anticommute. To preserve the correct commutation relations, this representation must be modified as follows:
\begin{equation}
    \begin{cases}
\hat{S}_j^{(z)}=\hat{c}_j^{\dag}\hat{c}_j-\frac{1}{2}
        \\[1.7ex]
\hat{S}_j^{(+)}=e^{\mathrm{i}\pi \sum\limits_{l<j}\hat{c}_l^{\dag}\hat{c}_l}\hat{c}_j^{\dag}=\hat{c}_j^{\dag}e^{-\mathrm{i}\pi \sum\limits_{l<j}\hat{c}_l^{\dag}\hat{c}_l}
        \\[1.7ex]
        \hat{S}_j^{(-)}=e^{\mathrm{i}\pi \sum\limits_{l<j}\hat{c}_l^{\dag}\hat{c}_l}\hat{c}_j=\hat{c}_je^{-\mathrm{i}\pi \sum\limits_{l<j}\hat{c}_l^{\dag}\hat{c}_l}.
    \end{cases}
\end{equation}
Regarding the symmetries of the two models, particle-number conservation in the fermionic representation, $\sum_i n_i$, corresponds to magnetization conservation in the spin representation, $\sum_i \sigma^z_i$.

\bibliography{bibliografia.bib}

@article{choi2017observation,
  title   = {Observation of discrete time-crystalline order in a disordered dipolar many-body system},
  author  = {Choi, Soonwon and Choi, Joonhee and Landig, Renate and Kucsko, Georg and Zhou, Hengyun and Isoya, Junichi and Jelezko, Fedor and Onoda, Shinobu and Sumiya, Hitoshi and Khemani, Vedika and von Keyserlingk, Curt and Yao, Norman Y. and Demler, Eugene and Lukin, Mikhail D.},
  journal = {Nature},
  volume  = {543},
  pages   = {221--225},
  year    = {2017},
  doi     = {10.1038/nature21426}
}

@article{else2017prethermal,
  title     = {Prethermal Phases of Matter Protected by Time-Translation Symmetry},
  author    = {Else, Dominic V. and Bauer, Bela and Nayak, Chetan},
  journal   = {Physical Review X},
  volume    = {7},
  number    = {1},
  pages     = {011026},
  year      = {2017},
  publisher = {American Physical Society},
  doi       = {10.1103/PhysRevX.7.011026}
}

@article{tang2026discrete,
  title     = {Discrete time crystals enabled by Floquet strong Hilbert space fragmentation},
  author    = {Tang, Ling-Zhi and Li, Xiao and Wang, Z. D. and Zhang, Dan-Wei},
  journal   = {Physical Review B},
  volume    = {113},
  number    = {21},
  pages     = {214303},
  year      = {2026},
  publisher = {American Physical Society},
  doi       = {10.1103/338m-mvsg}
}

@article{grazia3,
  title={Qubit-oscillator relationships in the open quantum Rabi model: the role of dissipation},
  author={Di Bello, G and Cangemi, L M  and Cataudella, V and De Filippis, G and Nocera, A and Perroni, C A},
  journal={The European Physical Journal Plus},
  volume={138},
  pages={135},
  year={2023},
  publisher={Società Italiana di Fisica (SIF)},
  doi = {10.1140/epjp/s13360-023-03714-x},
}

@article{grazia1,
  title={Local ergotropy and its fluctuations across a dissipative quantum phase transition},
  author={Di Bello, G and Farina, D and Jansen, D and  Perroni, C A and Cataudella, V and De Filippis, G},
  journal={Quantum Science and Technology},
  volume={10},
  pages={015049},
  year={2025},
  publisher={IOP},
  doi = {10.1088/2058-9565/ad9cbb},
}

@article{grazia2,
  title={Environment induced dynamical quantum phase transitions in two-qubit Rabi model},
  author={Di Bello, G and Ponticelli, A and 
  Pavan, F and Cataudella, V and De Filippis, G and de Candia, A and Perroni, C A},
  journal={Communications Physics},
  volume={7},
  pages={364},
  year={2024},
  publisher={Nature Publishing Group UK},
  doi = {10.1038/s42005-024-01855-8},
}

@article{wilczek2012,
  title={Quantum time crystals},
  author={Wilczek, Frank},
  journal={Physical review letters},
  volume={109},
  number={16},
  pages={160401},
  year={2012},
  publisher={APS},
  doi = {10.1103/PhysRevLett.109.160401},
}

@article{yao2018,
  title={Time crystals in periodically driven systems},
  author={Yao, Norman Y and Nayak, Chetan},
  journal={Physics Today},
  volume={71},
  number={9},
  pages={40--47},
  year={2018},
  publisher={American Institute of Physics},
  doi = {10.1063/PT.3.4020},
}

@article{khemani2019brief,
  title={A brief history of time crystals},
  author={Khemani, Vedika and Moessner, Roderich and Sondhi, SL},
  journal={arXiv preprint arXiv:1910.10745},
  year={2019},
  doi = {10.48550/arXiv.1910.10745},
}

@article{abanin2015periodically,
  title={Periodically driven ergodic and many-body localized quantum systems},
  author={Ponte, Pedro and Chandran, Anushya and Papi{\'c}, Z and Abanin, Dmitry A},
  journal={Annals of Physics},
  volume={353},
  pages={196--204},
  year={2015},
  publisher={Elsevier},
  doi = {10.1016/j.aop.2014.11.008},
}

@article{else2020discrete,
  title={Discrete time crystals},
  author={Else, Dominic V and Monroe, Christopher and Nayak, Chetan and Yao, Norman Y},
  journal={Annual Review of Condensed Matter Physics},
  volume={11},
  number={1},
  pages={467--499},
  year={2020},
  publisher={Annual Reviews},
  doi = {10.1146/annurev-conmatphys-031119-050658},
}

@article{formicola2025local,
  title = {Local ergotropy dynamically witnesses many-body localized phases},
  author = {Formicola, F. and Di Bello, G. and De Filippis, G. and Cataudella, V. and Farina, D. and Perroni, C. A.},
  journal = {Phys. Rev. Res.},
  volume = {7},
  issue = {4},
  pages = {043086},
  numpages = {11},
  year = {2025},
  month = {Oct},
  publisher = {American Physical Society},
  doi = {10.1103/2z1g-rgr9},
  url = {https://link.aps.org/doi/10.1103/2z1g-rgr9}
}

@article{serbyn2014quantum,
  title={Quantum quenches in the many-body localized phase},
  author={Serbyn, Maksym and Papi{\'c}, Zlatko and Abanin, Dmitry A},
  journal={Physical Review B},
  volume={90},
  number={17},
  pages={174302},
  year={2014},
  publisher={APS},
  doi = {10.1103/PhysRevB.90.174302},
}

@article{switzer2026realization,
  title={Realization of two-dimensional discrete time crystals with anisotropic Heisenberg coupling},
  author={Switzer, Eric D and Robertson, Niall F and Keenan, Nathan and Rodr{\'\i}guez-Alcaraz, {\'A}ngel and D’Urbano, Andrea and Pokharel, Bibek and Rahman, Talat S and Shtanko, Oles and Zhuk, Sergiy and Lorente, Nicol{\'a}s},
  journal={Nature Communications},
  volume={17},
  number={1},
  pages={605},
  year={2026},
  publisher={Nature Publishing Group UK London},
  doi = {10.1038/s41467-025-67787-1},
}

@article{barnes2019stabilization,
  title={Stabilization and manipulation of multispin states in quantum-dot time crystals with Heisenberg interactions},
  author={Barnes, Edwin and Nichol, John M and Economou, Sophia E},
  journal={Physical Review B},
  volume={99},
  number={3},
  pages={035311},
  year={2019},
  publisher={APS},
  doi = {10.1103/PhysRevB.99.035311},
}

@article{qiao2021floquet,
  title={Floquet-enhanced spin swaps},
  author={Qiao, Haifeng and Kandel, Yadav P and Dyke, John S Van and Fallahi, Saeed and Gardner, Geoffrey C and Manfra, Michael J and Barnes, Edwin and Nichol, John M},
  journal={Nature communications},
  volume={12},
  number={1},
  pages={2142},
  year={2021},
  publisher={Nature Publishing Group UK London},
  doi = {10.1038/s41467-021-22415-6},
}

@article{van2021protecting,
  title={Protecting quantum information in quantum dot spin chains by driving exchange interactions periodically},
  author={Van Dyke, John S and Kandel, Yadav P and Qiao, Haifeng and Nichol, John M and Economou, Sophia E and Barnes, Edwin},
  journal={Physical Review B},
  volume={103},
  number={24},
  pages={245303},
  year={2021},
  publisher={APS},
  doi = {10.1103/PhysRevB.103.245303},
}

@article{frantzeskakis2023time,
  title={Time-crystalline behavior in central-spin models with Heisenberg interactions},
  author={Frantzeskakis, Rafail and Van Dyke, John and Zaporski, Leon and Gangloff, Dorian A and Le Gall, Claire and Atat{\"u}re, Mete and Economou, Sophia E and Barnes, Edwin},
  journal={Physical Review B},
  volume={108},
  number={7},
  pages={075302},
  year={2023},
  publisher={APS},
  doi = {10.1103/PhysRevB.108.075302},
}

@article{li2020discrete,
  title={Discrete time crystal in the gradient-field Heisenberg model},
  author={Li, Bikun and Van Dyke, John S and Warren, Ada and Economou, Sophia E and Barnes, Edwin},
  journal={Physical Review B},
  volume={101},
  number={11},
  pages={115303},
  year={2020},
  publisher={APS},
  doi = {10.1103/PhysRevB.101.115303},
}

@article{chen2025discrete,
  title={Discrete time crystal and perfect many-body tunneling in a periodically driven Heisenberg spin chain},
  author={Chen, Xiaotong and Wu, Jianda},
  journal={arXiv preprint arXiv:2507.15565},
  year={2025},
  doi = {10.48550/arXiv.2507.15565},
}

@article{perez2006matrix,
  title={Matrix product state representations},
  author={Perez-Garcia, David and Verstraete, Frank and Wolf, Michael M and Cirac, J Ignacio},
  journal={arXiv preprint quant-ph/0608197},
  year={2006},
  doi = {10.26421/QIC7.5-6-1},
}

@article{white1992density,
  title={Density matrix formulation for quantum renormalization groups},
  author={White, Steven R},
  journal={Physical review letters},
  volume={69},
  number={19},
  pages={2863},
  year={1992},
  publisher={APS},
  doi = {10.1103/PhysRevLett.69.2863},
}

@article{schollwock2011density,
  title={The density-matrix renormalization group in the age of matrix product states},
  author={Schollw{\"o}ck, Ulrich},
  journal={Annals of physics},
  volume={326},
  number={1},
  pages={96--192},
  year={2011},
  publisher={Elsevier},
  doi = {10.1103/RevModPhys.77.259},
}

@article{fishman2022itensor,
	title={{The ITensor Software Library for Tensor Network Calculations}},
	author={Matthew Fishman and Steven R. White and E. Miles Stoudenmire},
	journal={SciPost Phys. Codebases},
	pages={4},
	year={2022},
	publisher={SciPost},
	doi={10.21468/SciPostPhysCodeb.4},
	url={https://scipost.org/10.21468/SciPostPhysCodeb.4},
}

@article{haegeman2016unifying,
  title={Unifying time evolution and optimization with matrix product states},
  author={Haegeman, Jutho and Lubich, Christian and Oseledets, Ivan and Vandereycken, Bart and Verstraete, Frank},
  journal={Phys. Rev. B},
  volume={94},
  number={16},
  pages={165116},
  year={2016},
  publisher={APS},
  doi = {10.1103/PhysRevB.94.165116},
}

@article{haegeman2011time,
  title={Time-dependent variational principle for quantum lattices},
  author={Haegeman, Jutho and Cirac, J Ignacio and Osborne, Tobias J and Pi{\v{z}}orn, Iztok and Verschelde, Henri and Verstraete, Frank},
  journal={Phys. Rev. Lett.},
  volume={107},
  number={7},
  pages={070601},
  year={2011},
  publisher={APS},
  doi = {10.1103/PhysRevLett.107.070601},
}

@article{hauke2016measuring,
  title={Measuring multipartite entanglement through dynamic susceptibilities},
  author={Hauke, Philipp and Heyl, Markus and Tagliacozzo, Luca and Zoller, Peter},
  journal={Nature Physics},
  volume={12},
  number={8},
  pages={778--782},
  year={2016},
  publisher={Nature Publishing Group UK London},
  doi = {10.1038/nphys3700},
}

@article{castellano2024extended,
  title={Extended local ergotropy},
  author={Castellano, Riccardo and Farina, Donato and Giovannetti, Vittorio and Acin, Antonio},
  journal={Physical Review Letters},
  volume={133},
  number={15},
  pages={150402},
  year={2024},
  publisher={APS},
  doi = {10.1103/PhysRevLett.133.150402},
}

@article{parlato2025quantum,
  title={Quantum Fisher information as a witness of non-Markovianity and criticality in the spin-boson model},
  author={Parlato, Daniele and Di Bello, Grazia and Pavan, Fabrizio and De Filippis, Giulio and Perroni, Carmine Antonio},
  journal={Physical Review B},
  volume={112},
  number={22},
  pages={224314},
  year={2025},
  publisher={APS},
  doi = {10.1103/fq4l-8v5g},
}

@article{zhang2017observation,
  title={Observation of a discrete time crystal},
  author={Zhang, Jiehang and Hess, Paul W and Kyprianidis, A and Becker, Petra and Lee, A and Smith, J and Pagano, Gaetano and Potirniche, I-D and Potter, Andrew C and Vishwanath, Ashvin and others},
  journal={Nature},
  volume={543},
  number={7644},
  pages={217--220},
  year={2017},
  publisher={Nature Publishing Group UK London},
  doi = {10.1038/nature21413},
}

@article{mi2022time,
  title={Time-crystalline eigenstate order on a quantum processor},
  author={Mi, Xiao and Ippoliti, Matteo and Quintana, Chris and Greene, Ami and Chen, Zijun and Gross, Jonathan and Arute, Frank and Arya, Kunal and Atalaya, Juan and Babbush, Ryan and others},
  journal={Nature},
  volume={601},
  number={7894},
  pages={531--536},
  year={2022},
  publisher={Nature Publishing Group UK London},
  doi = {10.1038/s41586-021-04257-w},
}

@article{von2016absolute,
  title={Absolute stability and spatiotemporal long-range order in Floquet systems},
  author={von Keyserlingk, Curt W and Khemani, Vedika and Sondhi, Shivaji L},
  journal={Physical Review B},
  volume={94},
  number={8},
  pages={085112},
  year={2016},
  publisher={APS},
  doi = {10.1103/PhysRevB.94.085112},
}

@article{frey2022realization,
  title={Realization of a discrete time crystal on 57 qubits of a quantum computer},
  author={Frey, Philipp and Rachel, Stephan},
  journal={Science advances},
  volume={8},
  number={9},
  pages={eabm7652},
  year={2022},
  publisher={American Association for the Advancement of Science},
  doi = {10.1126/sciadv.abm7652},
}

@article{gogolin2016equilibration,
  title={Equilibration, thermalisation, and the emergence of statistical mechanics in closed quantum systems},
  author={Gogolin, Christian and Eisert, Jens},
  journal={Reports on Progress in Physics},
  volume={79},
  number={5},
  pages={056001},
  year={2016},
  publisher={IOP Publishing},
  doi = {10.1088/0034-4885/79/5/056001},
}

@article{d2016quantum,
  title={From quantum chaos and eigenstate thermalization to statistical mechanics and thermodynamics},
  author={D'Alessio, Luca and Kafri, Yariv and Polkovnikov, Anatoli and Rigol, Marcos},
  journal={Advances in Physics},
  volume={65},
  number={3},
  pages={239--362},
  year={2016},
  publisher={Taylor \& Francis},
  doi = {10.1080/00018732.2016.1198134},
}

@article{Abanin,
  title={Many-body localization, thermalization, and entanglement},
  author={Abanin, D A and Altman, E and Bloch, I and Serbyn, M},
  journal={Rev. Mod. Phys.},
  volume={91},
  number={},
  pages={021001},
  year={2019},
  publisher={American Physical Society},
  doi = {10.1103/RevModPhys.91.021001},
}

@article{Nandki,
  title={Many-Body Localization and Thermalization in Quantum Statistical Mechanics},
  author={Nandkishore, R and Huse, D A},
  journal={Annu. Rev. Condens. Matter Phys.},
  volume={6},
  number={},
  pages={15-38},
  year={2015},
  publisher={Annual Reviews},
  doi = {10.1146/annurev-conmatphys-031214-014726},
}

@article{Alet,
  title={Many-body localization: An introduction and selected topics},
  author={Alet, F and Laflorencie, N},
  journal={C. R. Phys.},
  volume={19},
  number={},
  pages={498-525},
  year={2018},
  publisher={Elsevier},
  doi = {10.1016/j.crhy.2018.03.003},
}

@article{Sierant,
  title={Many-body localization in the age of classical computing},
  author={Sierant, P and  Lewenstein, M and Scardicchio, A and Vidmar, L and Zakrzewski, J},
  journal={Rep. Prog. Phys.},
  volume={88},
  number={},
  pages={026502},
  year={2025},
  publisher={IOP},
  doi = {10.1088/1361-6633/ad9756},
}

@article{ponte2015many,
  title={Many-body localization in periodically driven systems},
  author={Ponte, Pedro and Papi{\'c}, Zlatko and Huveneers, Fran{\c{c}}ois and Abanin, Dmitry A},
  journal={Physical review letters},
  volume={114},
  number={14},
  pages={140401},
  year={2015},
  publisher={APS},
  doi = {10.1103/PhysRevLett.114.140401},
}

@article{serbyn2013universal,
  title={Universal slow growth of entanglement in interacting strongly disordered systems},
  author={Serbyn, Maksym and Papi{\'c}, Zlatko and Abanin, Dmitry A},
  journal={Phys. Rev. Lett.},
  volume={110},
  number={26},
  pages={260601},
  year={2013},
  publisher={APS},
  doi = {10.1103/PhysRevLett.110.260601},
}

@article{Bardarson,
  title={Unbounded growth of entanglement in models of many-body localization},
  author={Bardarson, Jens H and Pollmann, Frank and Moore, Joel E},
  journal={Phys. Rev. Lett.},
  volume={109},
  number={1},
  pages={017202},
  year={2012},
  publisher={APS},
  doi = {10.1103/PhysRevLett.109.017202},
}

@article{lukin2019probing,
  title={Probing entanglement in a many-body--localized system},
  author={Lukin, Alexander and Rispoli, Matthew and Schittko, Robert and Tai, M Eric and Kaufman, Adam M and Choi, Soonwon and Khemani, Vedika and L{\'e}onard, Julian and Greiner, Markus},
  journal={Sci.},
  volume={364},
  number={6437},
  pages={256--260},
  year={2019},
  publisher={American Association for the Advancement of Science},
  doi = {10.1126/science.aau0818},
}

@article{allahverdyan2004maximal,
  title={Maximal work extraction from finite quantum systems},
  author={Allahverdyan, Armen E and Balian, Roger and Nieuwenhuizen, Th M},
  journal={EPL (Europhysics Letters)},
  volume={67},
  number={4},
  pages={565--571},
  year={2004},
  doi = {10.1209/epl/i2004-10101-2},
}

@article{alicki2013entanglement,
  title={Entanglement boost for extractable work from ensembles of quantum batteries},
  author={Alicki, Robert and Fannes, Mark},
  journal={Phys. Rev. E},
  volume={87},
  number={4},
  pages={042123},
  year={2013},
  publisher={APS},
  doi = {10.1103/PhysRevE.87.042123},
}

@article{SalviaGiovannetti,
  title={Optimal local work extraction from bipartite quantum systems in the presence of {H}amiltonian couplings},
  author={Salvia, Raffaele and De Palma, Giacomo and Giovannetti, Vittorio},
  journal={Phys. Rev. A},
  volume={107},
  number={1},
  pages={012405},
  year={2023},
  publisher={APS},
  doi = {10.1103/PhysRevA.107.012405},
}

@article{watanabe2015absence,
  title={Absence of quantum time crystals},
  author={Watanabe, Haruki and Oshikawa, Masaki},
  journal={Physical review letters},
  volume={114},
  number={25},
  pages={251603},
  year={2015},
  publisher={APS},
  doi = {10.1103/PhysRevLett.114.251603},
}

@book{strocchi2020symmetry,
  title={Symmetry breaking},
  author={Strocchi, Franco},
  year={2020},
  publisher={Springer},
  doi = {10.1007/978-3-540-73593-9},
}

@article{khemani2016phase,
  title={Phase structure of driven quantum systems},
  author={Khemani, Vedika and Lazarides, Achilleas and Moessner, Roderich and Sondhi, Shivaji L},
  journal={Physical review letters},
  volume={116},
  number={25},
  pages={250401},
  year={2016},
  publisher={APS},
  doi = {10.1103/PhysRevLett.116.250401}
}

@article{else2016floquet,
  title={Floquet time crystals},
  author={Else, Dominic V and Bauer, Bela and Nayak, Chetan},
  journal={Physical review letters},
  volume={117},
  number={9},
  pages={090402},
  year={2016},
  publisher={APS},
  doi = {10.1103/PhysRevLett.117.090402},
}

@article{yao2017discrete,
  title={Discrete time crystals: Rigidity, criticality, and realizations},
  author={Yao, Norman Y and Potter, Andrew C and Potirniche, I-D and Vishwanath, Ashvin},
  journal={Physical review letters},
  volume={118},
  number={3},
  pages={030401},
  year={2017},
  publisher={APS},
  doi = {10.1103/PhysRevLett.118.030401},
}

@article{li2025prethermal,
  title     = {Discrete time crystals in one-dimensional classical Floquet systems with nearest-neighbor interactions},
  author    = {Li, Zhuo-Yi and Zhang, Yu-Ran},
  journal   = {Physical Review B},
  volume    = {112},
  number    = {13},
  pages     = {134313},
  year      = {2025},
  publisher = {American Physical Society},
  doi       = {10.1103/7h9c-2n8w}
}

@article{sambe1973steady,
  title={Steady states and quasienergies of a quantum-mechanical system in an oscillating field},
  author={Sambe, Hideo},
  journal={Physical Review A},
  volume={7},
  number={6},
  pages={2203},
  year={1973},
  publisher={APS},
  doi = {10.1103/PhysRevA.7.2203},
}

@article{shirley1965solution,
  title={Solution of the Schr{\"o}dinger equation with a Hamiltonian periodic in time},
  author={Shirley, Jon H},
  journal={Physical Review},
  volume={138},
  number={4B},
  pages={B979},
  year={1965},
  publisher={APS},
  doi = {10.1103/PhysRev.138.B979},
}

@article{zel1967quasienergy,
  title={The quasienergy of a quantum-mechanical system subjected to a periodic action},
  author={Zel'Dovich, Ya B},
  journal={Sov. Phys. JETP},
  volume={24},
  number={5},
  year={1967},
}

@article{tsuji2023floquet,
  title={Floquet states},
  author={Tsuji, Naoto},
  journal={arXiv preprint arXiv:2301.12676},
  year={2023},
  doi = {10.1016/B978-0-323-90800-9.00241-9},
}

@article{lazarides2014periodic,
  title={Periodic thermodynamics of isolated quantum systems},
  author={Lazarides, Achilleas and Das, Arnab and Moessner, Roderich},
  journal={Physical review letters},
  volume={112},
  number={15},
  pages={150401},
  year={2014},
  publisher={APS},
  doi = {10.1103/PhysRevLett.112.150401},
}

@article{d2014long,
  title={Long-time behavior of isolated periodically driven interacting lattice systems},
  author={D’Alessio, Luca and Rigol, Marcos},
  journal={Physical Review X},
  volume={4},
  number={4},
  pages={041048},
  year={2014},
  publisher={APS},
  doi = {10.1103/PhysRevX.4.041048},
}

@article{nagao2026probing,
  title={Probing many-body localization crossover in quasiperiodic Floquet circuits on a quantum processor},
  author={Nagao, Kazuma and Shirakawa, Tomonori and Sun, Rongyang and Prelov{\v{s}}ek, Peter and Yunoki, Seiji},
  journal={arXiv preprint arXiv:2603.12675},
  year={2026},
  doi={10.48550/arXiv.2603.12675},
}

@article{de2019algebraic,
  title={Algebraic many-body localization and its implications on information propagation},
  author={De Tomasi, Giuseppe},
  journal={Physical Review B},
  volume={99},
  number={5},
  pages={054204},
  year={2019},
  publisher={APS},
  doi={10.1103/PhysRevB.99.054204}
}

@article{liu2018quantum,
  title={Quantum Fisher information and the localization properties of two interacting particles in one-dimensional systems},
  author={Liu, XM and Gao, GJ and Zhang, YM and Liu, J-M},
  journal={Solid State Communications},
  volume={279},
  pages={12--16},
  year={2018},
  publisher={Elsevier},
  doi={}
}

@article{liu2020quantum,
  title={Quantum Fisher information matrix and multiparameter estimation},
  author={Liu, Jing and Yuan, Haidong and Lu, Xiao-Ming and Wang, Xiaoguang},
  journal={Journal of Physics A: Mathematical and Theoretical},
  volume={53},
  number={2},
  pages={023001},
  year={2020},
  publisher={IOP Publishing},
  doi = {10.1088/1751-8121/ab5d4d},
}

@article{helstrom1969quantum,
  title={Quantum detection and estimation theory},
  author={Helstrom, Carl W},
  journal={Journal of statistical physics},
  volume={1},
  number={2},
  pages={231--252},
  year={1969},
  publisher={Springer},
  doi = {10.1007/BF01007479},
}

@book{holevo2011probabilistic,
  title={Probabilistic and statistical aspects of quantum theory},
  author={Holevo, Alexander S},
  volume={1},
  year={2011},
  publisher={Springer Science \& Business Media},
  doi = {10.1007/978-88-7642-378-9},
}

@article{gammelmark2014fisher,
  title={Fisher information and the quantum Cram{\'e}r-Rao sensitivity limit of continuous measurements},
  author={Gammelmark, S{\o}ren and M{\o}lmer, Klaus},
  journal={Physical review letters},
  volume={112},
  number={17},
  pages={170401},
  year={2014},
  publisher={APS},
  doi = {10.1103/PhysRevLett.112.170401},
}

@article{alipour2014quantum,
  title={Quantum metrology in open systems: dissipative Cram{\'e}r-Rao bound},
  author={Alipour, S and Mehboudi, M and Rezakhani, AT},
  journal={Physical review letters},
  volume={112},
  number={12},
  pages={120405},
  year={2014},
  publisher={APS},
  doi = {10.1103/PhysRevLett.112.120405},
}

@article{kongkhambut2022observation,
  title={Observation of a continuous time crystal},
  author={Kongkhambut, Phatthamon and Skulte, Jim and Mathey, Ludwig and Cosme, Jayson G and Hemmerich, Andreas and Ke{\ss}ler, Hans},
  journal={Science},
  volume={377},
  number={6606},
  pages={670--673},
  year={2022},
  publisher={American Association for the Advancement of Science},
  doi = {10.1126/science.abo3382},
}

@article{wu2024dissipative,
  title={Dissipative time crystal in a strongly interacting Rydberg gas},
  author={Wu, Xiaoling and Wang, Zhuqing and Yang, Fan and Gao, Ruochen and Liang, Chao and Tey, Meng Khoon and Li, Xiangliang and Pohl, Thomas and You, Li},
  journal={Nature Physics},
  volume={20},
  number={9},
  pages={1389--1394},
  year={2024},
  publisher={Nature Publishing Group UK London},
  doi = {10.1038/s41567-024-02542-9},
}

@article{russo2025quantum,
  title={Quantum dissipative continuous time crystals},
  author={Russo, Felix and Pohl, Thomas},
  journal={Physical Review Letters},
  volume={135},
  number={11},
  pages={110404},
  year={2025},
  publisher={APS},
  doi = {10.1103/dc2s-94gv},
}

@article{wu2026dissipative,
  title={Dissipative time crystal in a thermal Rydberg gas based on microwave dressing},
  author={Wu, Lianglong and Xiao, Mengzhuo and Xu, Yinuo and Chen, Haixia and Wei, Dong},
  journal={Physical Review A},
  volume={113},
  number={5},
  pages={053116},
  year={2026},
  publisher={APS},
  doi = {10.1103/yb4y-lwzm},
}

@article{khasseh2026semiclassical,
  title={Semiclassical Langevin dynamics of long-range dissipative time crystals},
  author={Khasseh, Reyhaneh and Fazio, Rosario and Russomanno, Angelo},
  journal={arXiv preprint arXiv:2607.03486},
  doi = {10.48550/arXiv.2607.03486},
  }

\end{document}